\documentclass[reprint,pra,twocolumn,amsmath,amssymb,superscriptaddress,nofootinbib,showpacs,aps]{revtex4-1}

\pdfoutput=1
\usepackage{graphicx}
\usepackage{bm}
\usepackage{times}

\usepackage{color}
\definecolor{blue}{RGB}{0,0,255}
\usepackage[colorlinks,citecolor=blue,linkcolor=blue,urlcolor=blue]{hyperref}

\begin{document}

\preprint{APS/123-QED}

\def\BE{\begin{equation}}
\def\EE{\end{equation}}
\def\BY{\begin{eqnarray}}
\def\EY{\end{eqnarray}}
\def\BI{\begin{itemize}}
\def\EI{\end{itemize}}
\def\L{\label}
\def\nn{\nonumber}
\def\({\left (}
\def\){\right)}
\def\[{\left [}
\def\]{\right]}
\def\<{\langle}
\def\>{\rangle}
\def\o{\overline}
\def\BA{\begin{array}}
\def\EA{\end{array}}
\def\dsp{\displaystyle}
\def\ds{\displaystyle}
\def\k{\kappa}
\def\dd{\delta}
\def\D{\Delta}
\def\w{\omega}
\def\W{\Omega}
\def\a{\alpha}
\def\b{\beta}
\def\d{\partial}
\def\g{\gamma}
\def\tt{\theta}
\def\t{\tau}
\def\s{\sigma}
\def\+{\dag}
\def\x{\xi}
\def\m{\mu}
\def\l{\lambda}
\def\ii{\textbf{i}}
\def\={\approx}
\def\xc{\frac{2x}{c}}
\def\->{\rightarrow}
\def\r{\vec{r}}
\def\k{\vec{k}}
\def\sinc{\mathrm{sinc}}
\def\xx{\textbf{x}}
\def\yy{\textbf{y}}
\def\qq{\textbf{q}}
\def\rr{\boldsymbol{\rho}}
\def\length{0.45}
\def\-{\overline}
\newcommand{\ud}{\,\mathrm{d}} 

\title{Non-linear photon subtraction from a multimode quantum field}

\author{Valentin A. Averchenko}
\email{valentin.averchenko@lkb.upmc.fr}
\affiliation{Laboratoire Kastler Brossel, Universit\'{e} Pierre et Marie Curie-Paris 6, ENS, CNRS; 4 place Jussieu, 75252 Paris, France}

\author{Val\'erian Thiel}
\affiliation{Laboratoire Kastler Brossel, Universit\'{e} Pierre et Marie Curie-Paris 6, ENS, CNRS; 4 place Jussieu, 75252 Paris, France}

\author{Nicolas Treps}
\affiliation{Laboratoire Kastler Brossel, Universit\'{e} Pierre et Marie Curie-Paris 6, ENS, CNRS; 4 place Jussieu, 75252 Paris, France}

\date{\today}

\begin{abstract}
We consider theoretically how to extract mode dependent single photons from a time/frequency multimode non-classical beam. To achieve this task, we calculate the properties of sum frequency generation with a pulse shaped pump, taking into account both temporal and spatial degree of freedom. We show that using a non-collinear configuration it is possible to achieve a mode dependent weakly reflective beam splitter, with Schmidt number compatible with photon extraction for continuous variable regime tasks. We explicit the possible application to the degaussification of highly multimode squeezed frequency combs.
\end{abstract}

\pacs{42.50.-p, 42.65.Ky, 42.50.Dv, 03.65.Ud}

\maketitle

\section{Introduction}

In order to design experimentally viable quantum information processing systems, several key technological achievements are required. One of the most important is obtaining a scalable multimode device. To this aim, many systems have been proposed, being either atomic \cite{Lewenstein2007, *HAFFNER2008, *Bloch2012}, solid state \cite{Fowler2012a, *Barends2014} or optics-based \cite{Walther2005, *Aoki2009, *Ukai2011, *Yokoyama2013}. In particular, the latter approach has recently offered a highly promising concept that consists in using the intrinsic multimode nature of a single beam of light to encode and manipulate quantum information \cite{Pysher2011,Armstrong2012}. Either q-bits, where single photons carry the information, or q-modes, where the field quadratures are used as quantum carriers, have been considered. Even if both approaches involve very different concepts both in information encoding and measurement strategies, the optical technics at play are very similar.

We consider here the use of the light frequency degree of freedom to encode information. Recently, several approaches have led to the generation of highly multimode light \cite{Pysher2011, Roslund2013} in that regime. However, one of the key operations towards quantum information processing remains to be performed: the ability to arbitrarily manipulate frequency modes.

Toward that goal, particular attention has been paid to nonlinear up-conversion as a tool to manipulate frequency modes while preserving their quantum states.
For example, it has been used to up-convert the frequency of a single photon preserving non-classical correlations with a reference photon \cite{Tanzilli2005} or to up-convert a squeezed vacuum state \cite{Vollmer2014}.
The control of a strong coherent field mediating the process provides a way to coherently manipulate spectral properties of an up-converted light \cite{Kielpinski2011}. In \cite{Lavoie2013a} spectral compression of an up-converted photon has been performed.
The group of C. Silberhorn \cite{Eckstein2011} has proposed to use sum-frequency generation in a waveguide to achieve a so-called quantum pulse gate able to extract a given frequency mode depending on a gate beam.
It has been shown that choosing a non-linear waveguide of proper length and dispersion will completely up-convert one single mode of the signal field, giving access to its parameters independently from the rest. In work \cite{Huang2013}, the sequential conversion of multiple modes has been put forward to perform mode-resolved measurements.

	\begin{figure}[t]
	\center{\includegraphics[width=0.85\linewidth]{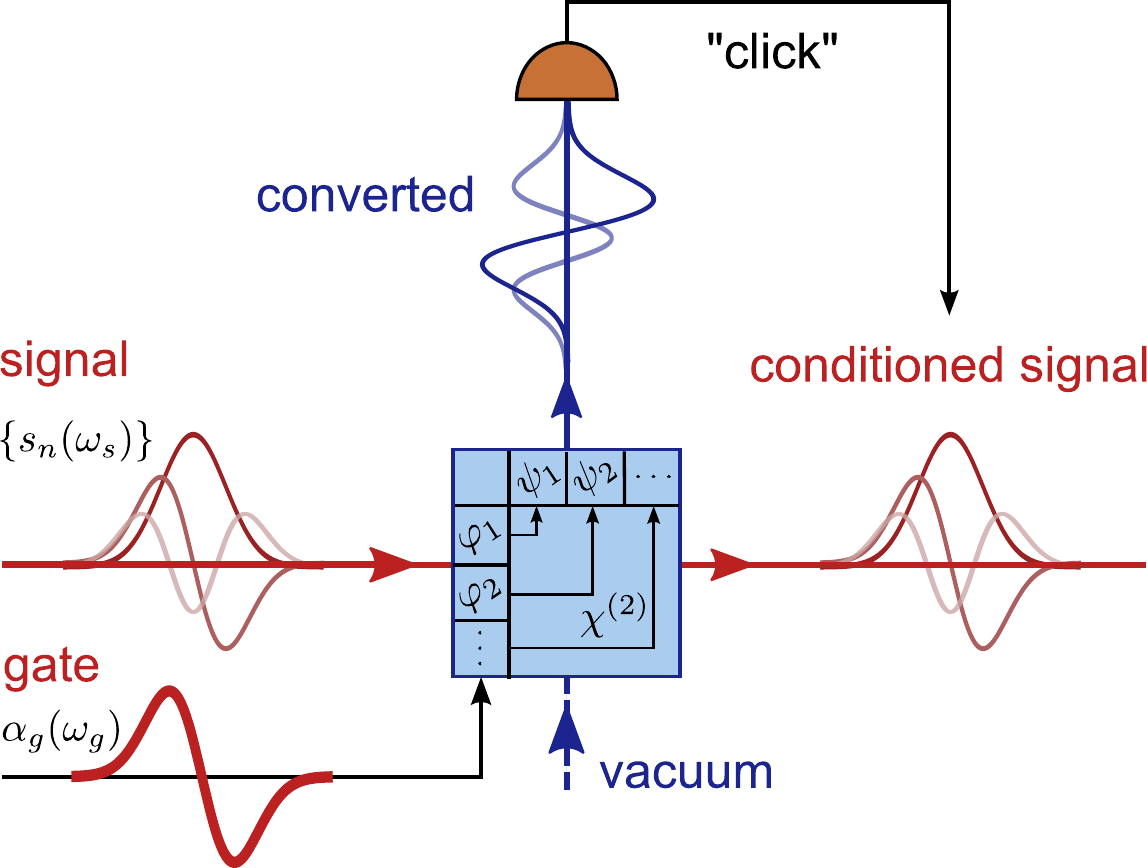}}
	\caption{Conditional photon subtraction from a signal field via parametric up-conversion process. The input signal field is composed of broadband quantum modes with spectral profiles $\{s_n(\w_s)\}$. The set $\{\varphi_m(\w_s)\}$ defines input modes of a photon subtraction arrangement. A subtracted photon is up-converted into the corresponding output mode $\{\psi_m(\w_c)\}$. The modes are controlled by the gate field with a spectral distribution $\a_g(\w_g)$.}
	\label{fig:QPG}
	\end{figure}

We propose here to use nonlinear up-conversion for continuous variable tasks such as photon subtraction for the generation of non-gaussian states. These tasks are to be performed on a multimode quantum state encoded on the frequency degree of freedom of a single beam, which we will call the signal field. The general principle of the non-linear process is outlined in Fig. \ref{fig:QPG}. The signal field and a strong coherent field undergo a weak parametric interaction in a non-linear medium that leads to the probabilistic up-conversion of a photon from the signal field. The selective detection of this  up-converted light leads to the conditional photon subtraction from the signal field.
Therefore, the up-conversion process plays the role of a low-reflectivity beam-splitter like in the conventional photon subtraction scheme \cite{Dakna1996}. However in the present case, the process is intrinsically multimode \cite{Eckstein2011} and the spectral properties of photon extraction depend both on the non-linear medium properties and the gate field spectral amplitude. In turn, this influences the purity and the quantum state of the conditioned state, which can be tuned by acting on experimental parameters. This proposed method is suited to the subtraction of a broadband photon and is an alternative to the standard photon subtraction scheme that uses a beam-splitter together with a narrow-band spectral filter \cite{Wenger2004}.

We develop in this paper the theory of  non-collinear sum-frequency generation in a  crystal taking into account both temporal and spatial degrees of freedom. We consider the example of the degaussification of multimode squeezed optical frequency combs, as those produced experimentally in \cite{Roslund2013}. These combs represent a promising resource for quantum computing \cite{Ferrini2013} and optical metrology \cite{Lamine2008}.

This paper is organized as follows:
in Sec. \ref{Model}, we present the theory of non-collinear sum-frequency generation between a signal and a gate field.
In Sec. \ref{State} we calculate the state of the signal field conditioned on the detection of an up-converted photon.
In Sec. \ref{Apprx} we calculate the analytical conditions for the single-mode regime of a photon subtraction.
In Sec. \ref{Num} we present some numerical results, such as the number of up-converted modes and their spectral profiles or the purity of a conditioned state, for parameters that are relevant for squeezed frequency combs.
Sec.  \ref{Concl} concludes the paper.

\section{Photon subtraction via parametric up-conversion}\L{Model}
	\begin{figure}
	\center{\includegraphics[width=0.9\linewidth]{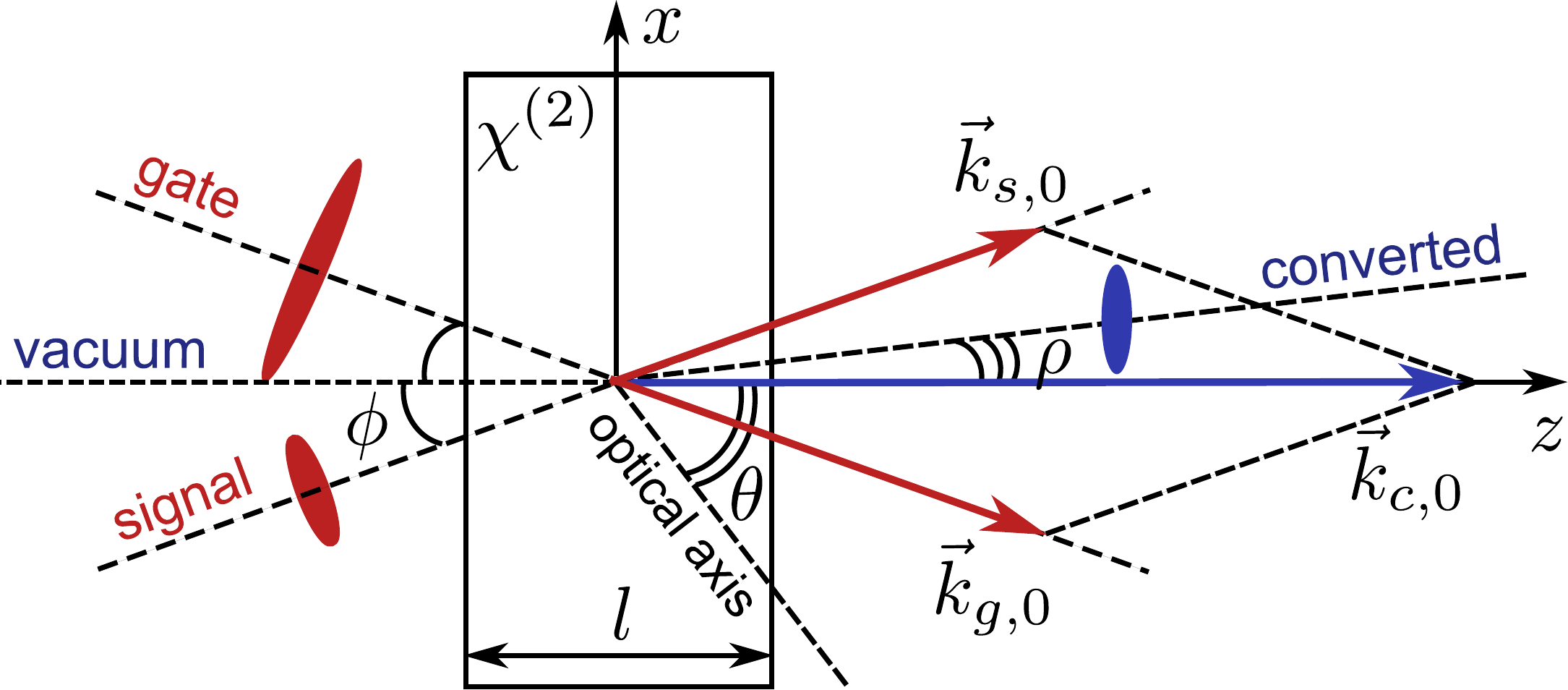}}
	\caption{Non-collinear sum-frequency generation. $\phi$ is a non-collinear mix-angle; $\theta$ is a phase-matching angle for type-I frequency degenerate interaction; $\rho$ is a birefringent walk-off angle of up-converted light with the extraordinary polarization.
	}
	\label{fig:non-coll_coord}
	\end{figure}
The proposed nonlinear photon subtraction is based on sum-frequency generation between a weak multimode quantum field (signal), and an intense reference field (gate).
As a possible implementation, we consider the non-collinear arrangement presented in Fig. \ref{fig:non-coll_coord}. In this configuration, the up-converted photon is automatically separated from the rest, thus type-I interaction where signal and gate fields are linearly polarized along the ordinary axis of the crystal can be used. In that case group velocities are matched independently of the non-collinear angle between the fields.
Then, in analogy with the parametric down-conversion process \cite{U'Ren2006}, one may expect the single-mode regime to be achieved with a long crystal.
Additional details of the configuration are presented in Appendix \ref{Exp}.

Consider now that the system input quantum state is expressed as the product of a given state of the signal field (e.g multimode squeezed light from \cite{Roslund2013}) and a vacuum state of the frequency doubled field (the converted field):
	\begin{align}
	& |\textrm{in}\> = |\textrm{in}_s\> \otimes |\textrm{vac}_c\> \L{in}
	\end{align}
The gate field is assumed to be intense and unchanged by the weak interaction, and thus is modeled by a classical quantity.
For a weak parametric interaction between the fields, the resulting state in the first order perturbation theory reads $|\textrm{out}\> \approx |\textrm{in}\> + |\phi\>$.
The second term represents the up-conversion of a signal photon of frequency and transverse momentum $(\w_s,\qq_s)$ into a photon $(\w_c,\qq_c)$ (for details, see Appendix \ref{state})
	\begin{align}
	\nn |\phi\> = &C \int \ud\w_c \ud \qq_c \ud\w_s \ud\qq_s \; \\
	& \times L(\w_c, \qq_c, \w_s, \qq_s) \; \hat{a}_c^\dag(\w_c, \qq_c) \hat{a}_s(\w_s, \qq_s) |\textrm{in}\> \L{Hdt}
	\end{align}
As photon detection will be performed on the up-converted field, only this second term of the output wave function is relevant for further calculation. Here $C=\varepsilon_0 \chi^{(2)} {\cal E}_s {\cal E}_c l \sqrt{W_g}/((2\pi)^{3/2} i\hbar \sqrt{2\varepsilon_0 n_g c})$ is a constant involving scaling factors ${\cal E}_{j} = \sqrt{{\hbar \w_{j,0}}/{2 \varepsilon_0 n_{j} c}}$ for signal and converted fields ($j=c,s$); $n_{c,s,g}$ are refractive indices at the carrier frequencies $\w_{c,s,g}$; $l$ is the length of the crystal; $W_g$ represents the energy in a gate pulse.
The transfer function is a product of the normalized spatio-temporal profile of the gate pulse and the phase-matching function
	\begin{align}
	& L(\w_c, \qq_c, \w_s, \qq_s) = \a_g(\w_g,\qq_g) \; \sinc(\D k l/2) \L{L4}
	\end{align}
where the following conditions representing the conservation of energy and momentum have to be fulfilled:
	\begin{align}
	& \w_g + \w_s - \w_c = 0, \L{dw}\\
	& q^y_g + q^y_s - q^y_c =0 , \L{dp}\\
	& (q^x_g + q^x_s) \cos\phi - q^x_c + (-k_g+k_s) \sin\phi =0 , \L{dq}\\
	& (k_g + k_s) \cos\phi  - k_c + (q^x_g-q^x_s)\sin\phi + q^x_c \tan\rho = \D k \L{dk}
	\end{align}
These expressions are obtained when neglecting the effect of diffraction.
One sees that the conservation of the transverse momentum along the $x$-axis (\ref{dq}) involves longitudinal momenta depending on the frequency and vice versa (\ref{dk}).
Therefore up-conversion dynamics depends on the spatial profiles of the interacting pulses and one has to treat both the spatial and temporal degrees of freedom simultaneously.
It is worth stressing that this coupling appears purely due to the non-collinear configuration, just like in the case of non-collinear down-conversion \cite{Valencia2007} and does not involve diffraction effects, in contrast, for example, to the spatio-temporal X-entanglement \cite{Gatti2009}.
	\begin{figure}	\center{\includegraphics[width=0.75\linewidth]{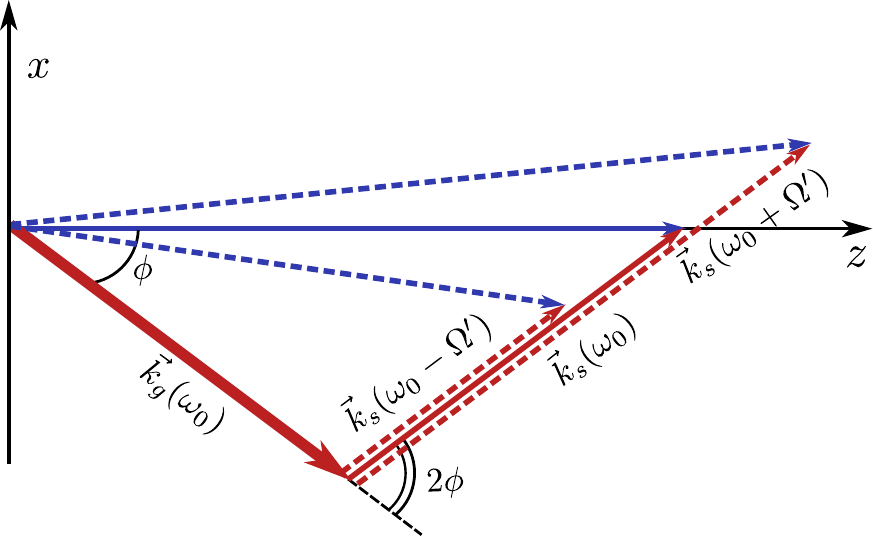}}

	\caption{A sketch of the mechanism leading to the spatio-temporal coupling in non-collinear  parametric up-conversion: three components of a signal field are parametrically scattered on a strong gate wave into non-collinear converted waves.}
	\label{fig:degrees_mix}
	\end{figure}
The physical mechanism underlying the coupling is depicted in Fig.~\ref{fig:degrees_mix}: three sample collinear spectral components of the signal field undergo up-conversion into waves propagating at different angles/having different transverse momenta.
In the collinear case limit where $\phi=0$ and under negligibly small spatial walk-off $\rho=0$, every up-converted waves are parallel and the transfer function (\ref{L4}) is factorized into spatial and temporal parts that may be treated independently.

For further analysis, let us apply the Schmidt decomposition of the transfer function (\ref{L4}). The decomposition reveals the underlying multimode properties of  the parametric process \cite{Law2004}. In the collinear case, the decomposition may be applied independently to the temporal and spatial parts of the factorisable transfer function. In the non-collinear case, the decomposition of the function is a more challenging task. The situation is simplified mathematically in the experimentally reasonable situation where the signal field carries photons that are only in a well-defined spatial mode.
For example, for a squeezed frequency comb generated within an optical cavity, the spatial mode, usually TEM$_{00}$, is imposed by the cavity.
Let the mode profile be described by the transverse momentum distribution $u_s(\qq_s)$.
The amplitude of the signal field in the expression (\ref{Hdt}) may then be represented as
	\begin{align}
	& \hat{a}_s(\w_s,\qq_s) = \hat{a}_{s}(\w_s) u_{s}(\qq_s) + \mbox{vacuum modes}
	\end{align}
Under a weak up-conversion process, the vacuum modes do not change their states and one can drop them out from further consideration.
In that case, the state of the non-vacuum signal mode and the up-converted field reads
	\begin{align}
	\nn |\phi\> = & \; C \int \ud\w_c \ud \qq_c \ud\w_s \\
	& \times L(\w_c, \qq_c, \w_s) \, \hat{a}_c^\dag(\w_c, \qq_c) \, \hat{a}_s(\w_s) |\textrm{in}\> \L{f3}
	\end{align}
In this expression, we defined the reduced transfer function
	\begin{align}
	\nn L(\w_c, \qq_c, \w_s) &\equiv \int \ud\qq_s L(\w_c, \qq_c, \w_s,\qq_s) u_s(\qq_s)\\
	&=\sum\limits_{m=1}^\infty \l_m \, \psi_m(\w_c, \qq_c) \, \varphi_m^*(\w_s)  \L{L4_decomp}
	\end{align}
The last equality represents the  Schmidt decomposition of the function into signal and up-converted parts. The real-valued coefficients of the decomposition can be ordered as $\l_1\geq\l_2\geq\l_3\geq \ldots \geq 0 $. The Schmidt functions $\left\{\psi_n\right\}$ and $\left\{\varphi_n\right\}$ are orthonormal sets.
Substituting the decomposition into (\ref{f3}) and defining broadband operators for the signal $\hat A_{s,m} = \int \ud\w_s \; \hat a_{s}(\w_s) \varphi_m^*(\w_s)$ and the converted $\hat A_{c,m} = \int \ud\w_c d \qq_c  \; \hat a_{c}(\w_c,\qq_c) \psi_m^*(\w_c,\qq_c)$ fields with the commutation condition $[\hat A_{j,m}, \hat A_{j,m'}^\dag]=\dd_{mm'}$, the state (\ref{f3}) takes the form
	\begin{align}
	& |\phi\> = C \sum\limits_{m=1}^{\infty} \l_m \hat A_{c,m}^\dag \hat A_{s,m} |\text{in}\> \L{wave_func_BS}
	\end{align}
The expression shows that a weak up-conversion process operates as a beam-splitter for broadband modes, subtracting a photon from the input Schmidt mode  $\varphi_m$ into the output mode $\psi_m$. The process goes in parallel with a probability for each mode that is proportional to $\lambda_m^2$. The number of modes involved may be efficiently characterized with the Schmidt number
	\begin{align}
	& K=\frac{(\sum\limits_m \l_m^2)^2}{\sum\limits_m \l_m^4}
	\end{align}
For the wave-guide configuration \cite{Eckstein2011}, it has been shown that the parametric up-conversion may be switched into a single-mode regime when all the Schmidt coefficients of the decomposition tend to zero except the first one $\l_{m>1} \rightarrow 0$, in which case $K \rightarrow 1$. In our arrangement, it means that the photon subtraction becomes mode selective. We will derive the sufficient conditions for this regime in Sec. \ref{Apprx} for a non-collinear free space configuration.

Both the modes and the Schmidt coefficients may be obtained in a straightforward manner by calculating the eigen-value decomposition of an auxiliary hermitian function $G(\w_s,\w'_s)$:
	\begin{align}
	\nn G(\w_s, \w_s') &\equiv \int \ud\w_c \ud \qq_c \, L^*(\w_c, \qq_c, \w_s) \, L(\w_c, \qq_c, \w'_s)\\
	& = \sum\limits_{m=1}^{\infty} \l_m^2 \, \varphi_m(\w_s) \, \varphi_m^*(\w'_s) \L{G_def}
	\end{align}
The eigen-values of this decomposition correspond to the squared Schmidt coefficients and the eigen-functions  to the subtraction modes.
We will employ this procedure for numerical calculations in Sec. \ref{Num}. In turn, the calculation of the up-converted modes is a more complicated task due to the spatio-temporal structure of the modes. However in the considered photon subtraction scheme (see Fig. \ref{fig:QPG}), the properties of the up-converted photons are not relevant. All the photons are collected with a non mode-selective photon-counting detector. Thus in the following, we will omit the detailed consideration of the up-converted modes, except in Sec. \ref{Apprx} where the analytical expressions are obtained in the single-mode regime.

\section{Conditioned state of the signal field}\L{State}

In this section we calculate the state of the signal field after conditioning on the up-converted photon. While we will give a general expression for the output density matrices, we will take specific examples for a multimode signal field state that can be factorized in a given mode basis (note that this is always the case for pure gaussian states \cite{Braunstein2005}). We name this eigen-mode basis $\{s_n(\w_s)\}$, each $s_n(\omega_s)$ describing the spectral amplitude of each signal mode.  The signal input state in that basis may then be written
\begin{align}\label{signalstate}
|\textrm{in}_s\> = \bigotimes_n |\textrm{in}_{s,n}\>.
\end{align}

We now consider the detection of the up-converted beam with a unity quantum efficiency single photon detector that is insensitive to the spatial and temporal profiles of the beam. The probability of a photon subtraction
is given by the norm of the wave-function (\ref{wave_func_BS})
	\begin{align}
	P= \<\phi|\phi\> =  |C|^2 \sum\limits_{m=1}^\infty \l_m^2 \; \<\textrm{in}_s| \hat{A}_{s,m}^\dag  \hat{A}_{s,m} |\textrm{in}_s\> \L{P1}
	\end{align}
and is equal to the sum of the up-conversion probabilities for each input mode multiplied by the corresponding number of photons in the mode.
Here we used the explicit expression for the initial state of the fields (\ref{in}).
In turn, the conditioned state of the signal field is described by the density  matrix
	\begin{align}
	\nn \hat \rho_s &=\frac{\mbox{Tr}_c\{|\phi\>\<\phi|\}}{P}\\&= \frac{|C|^2}{P}  \sum\limits_{m=1}^\infty \l_m^2 \; \hat{A}_{s,m} |\text{in}_s\> \<\text{in}_s| \hat{A}_{s,m}^\dag \L{p1}
	\end{align}
One can see that, in general, this state is mixed.
However, when the up-conversion process is single-mode (i.e. only $\lambda_1\neq 0$), the state is pure and reads
	\begin{align}
	|\textrm{out}_s\> \propto \hat{A}_{s,1} |\textrm{in}_s\>
	\end{align}
This expression means that a photon is subtracted from a broadband mode defined by parametric up-conversion.
In the most general case, there is no coincidence between the subtraction mode and any particular eigen-mode of the signal field of equation (\ref{signalstate}). Nevertheless, in principle this may be achieved by changing the profile of the gate pulses.
Then, for example, if the signal field is a multimode squeezed light, it brings the matched mode into a cat-like state \cite{Dakna1996} while the rest remains unchanged.
In the non-matched case, the photon is subtracted from a superposition of signal modes, giving rise to cat-like states superposition.

To calculate the purity of the conditioned state, let us denote $\{N_{s,n}\}$  the mean photon number in the signal field eigen-modes $\{s_n(\w_s)\}$. We define the overlap matrix $O_{mn} = \int \ud\w \; \varphi^*_m(\w_s) s_n(\w_s)$ between signal modes $\{s_n(\w_s)\}$ and subtraction modes $\{\varphi_m(\w_s)\}$.
One gets expressions for the subtraction probability (\ref{P1}) in the alternative form
	\begin{align}
	P = |C|^2 \sum_{m,n} \l_m^2 |O_{mn}|^2 N_{s,n} \L{P_multi}
	\end{align}
The purity (\ref{p1}) is then given by
	\begin{align}
    \mbox{Tr}(\hat{\rho}_s^2) = \frac{\sum\limits_{m,m'} \l_m^2 \l_{m'}^2 \left|\sum\limits_n O_{mn} O_{m'n}^* N_{s,n}\right|^2}{\(\sum\limits_{m,n} \l_m^2 |O_{mn}|^2 N_{s,n}\)^2} \L{purity}
	\end{align}
This state is pure in the single-mode regime when $\l_{m>1}=0$, as it was stressed above. Another possibility to get a pure state is when the initial state of the signal field is in a single-mode, i.e. when all modes are in a vacuum state except for one: $N_{s,n>1} = 0$.
An additional control over the conditioned state and its purity is achieved by tailoring subtraction modes and their overlap $O_{mn}$ with modes of the signal field.

\section{Single-mode regime of up-conversion: analytical treatment}\L{Apprx}

In order to calculate explicitly the subtraction modes and  their corresponding coefficients (\ref{L4_decomp}), let us make some additional assumptions.
Let us first assume that the spatio-spectral amplitude of the gate field is factorized
	\begin{align}
	& \a_g(\w_g,\qq_g) = \a_g(\w_g) u_g(\qq_g)
	\end{align}
where $\a_g(\w_g)$ is a spectral profile and $u_g(\qq_g)$ is the product of two identical functions along $x$ and $y$, i.e. $u_g(\qq_g) = u_g(q_g^x)u_g(q_g^y)$. The factorization means that we do not consider pulses with tilted fronts \cite{Torres2010}. In principles, front-tilt may provide an additional control over the system \cite{Torres2005a, Hendrych2007}.

We also assume that the transverse profile of the gate pulse is broader than the signal one, thus guaranteeing their overlap over the whole crystal length, as depicted on Fig. \ref{fig:non-coll_coord}. The gate beam is then approximated as a plane wave by putting $q^y_g, q^x_g=0$ in the expressions (\ref{dp}-\ref{dq}).
As a result of conservation of momentum in the $y$-direction (\ref{dp}), it follows that an up-converted photon and a signal photon share the same momentum $q^y_c = q^y_s$. Since this degree of freedom is uncoupled from the others, the distribution of momentum of up-converted photons in the $y$-direction coincides with the distribution of signal photons.
The problem is then reduced to a 2-D case by defining the following amplitude
	\begin{align}
	& \hat a_c(\w_c,q^x_c) = \int \ud q^y_c \; \hat a_c(\w_c,\qq_c) \; u_s(q^y_c)
	\end{align}
where we have also assumed that the signal beam is factorized, i.e. $u_s(\qq_s) = u_s(q^x_s) u_s(q^y_s)$. In the following we will omit the $x$ index.

An additional assumption concerns the dispersion properties of the crystal. Introducing the relative frequencies $\W_j = \w_j - \w_{j,0}$ for each field $(j=c,s,g)$, where $\w_{j,0}$ is the mean frequency of field $j$, we expand the wave-vectors to the first order $k_j \approx k_{j,0} + k'_j \; \W_j$.
We get the following expression for the wave-function (\ref{f3}):
	\begin{align}
	\nn |\phi\> = & C' \int \ud\W_c \ud q_c \ud\W_s \\
	&\times L(\W_c, q_c, \W_s) \; \hat{a}_c^\dag(\W_c, q_c) \hat{a}_s(\W_s)|\textrm{in}\>  \L{f_out}
	\end{align}
 where the proportionality coefficient is given by
	\begin{align}
	& C' = C \(\int \ud q_g u_g(q_g)\)^2
	\end{align}
The results of the previous section remain of course valid under the substitution $C \rightarrow C'$. The explicit expression of the transfer function then reads
	\begin{align}
	\nn L(& \W_c, q_c, \W_s) = \a_g(\W_c-\W_s)\\
	\nn & \times u_s\[ (\cos\phi)^{-1} \;q_c+k'_s \tan\phi \; (\W_c-2\W_s) \] \\
	\nn & \times \sinc\bigg\{\Big[(k'_c-k'_s \cos\phi) \W_c+k'_s \tan\phi \sin\phi (\W_c-2\W_s) \\
	& \qquad\qquad + (\tan\phi-\tan\rho) q_c\Big] l/2\bigg\} \L{LWqW}
	\end{align}
Here we used the phase-matching condition for the type-I degenerate process: $(k_{g,0} + k_{s,0})\cos\phi = k_{c,0}$  and $k_{s,0} = k_{g,0}$. Furthermore gate and signal pulses, ordinary polarized, have equal group velocities $k'_g=k'_s$.

Using a Gaussian approximation for the transfer function (\ref{LWqW}), let us estimate analytically in this section the number of subtraction modes (i.e. the Schmidt number) and how single-mode regime can be obtained.
For this purpose, we approximate the sinc function by: $\mathrm{sinc}(x) \approx e^{-\g x^2} , \g =0.193$. The numerical value of $\gamma$ is chosen so that both functions exhibit the same full width at half maximum.
We model the transverse profiles of the signal and gate beams as Gaussian beams with widths $w_{s,g}$
	\begin{align}
	& u_{j}(q_j) = \frac{\sqrt{w_{j}}}{\pi^{1/4}} \exp(-w_{j}^2 q_j^2/2), \qquad j=s,g \L{u}
	\end{align}
Considering a gate field with a much wider transverse profile, we choose $w_g \gg w_s$.
We also consider the specific case where the gate pulses have a Gaussian spectral profile
	\begin{align}
	& \a_g(\W_g) = \frac{\sqrt{\t_g}}{\pi^{1/4}} \exp(-\t_g^2 \W_g^2/2) \L{a_g}.
	\end{align}
Assuming small non-collinear and walk-off angles, we expand the trigonometric functions involved into the expression (\ref{LWqW}) to the lowest order. Under these condition, we also neglect the group velocity birefringence by approximating the group velocity of the up-converted field with its value for the collinear case, i.e. $k_c' \approx k_c'|_{\phi=0}$.
Using the procedure presented in Appendix \ref{Schmidt-calc}, one gets an expression for the Schmidt number that may be written as the following function of the crystal length and signal beam focusing
	\begin{align}
	& K(l,w_s) = \sqrt{(a(l) w_s^2+b)(c(l) w_s^{-2}+d)} \L{Klw}
	\end{align}
where $a(l) = 1+{2\t_g^2}/{\g (k_c'-k_s')^2 l^2}$, $b=(\phi-\rho)^2 {\t_g^2}/{(k_c'-k_s')^2}$, $c(l)=1+2\phi^4 \g k_s'^2 l^2/\t_g^2$ and $d=4\phi^2 {k_s'^2}/{\t_g^2}$.
One can show that the Schmidt number has a minimum value
	\begin{align}
	& K_{\text{min}} = 1 + \frac{\phi^2+|\phi (\phi-\rho)|}{\phi_0^2} \L{Kmin}
	\end{align}
under the optimal values of the crystal length and signal beam focusing
	\begin{align}
	& l_{\text{opt}} = \frac{\t_g}{k_s'} \frac{1}{\sqrt{2\gamma} \, \phi_0 \, |\phi|}, \L{lopt}\\
	& w_{\text{opt}} =  \frac{\t_g}{k_s'} \frac{\sqrt{\left|\rho/\phi-1\right|}}{2\phi_0} \L{wopt}
	\end{align}
Here we also defined a characteristic  non-collinear angle
	\begin{align}
	& \phi_0 = \sqrt{(k_c'/k_s'-1)/2} \L{f0}
	\end{align}
In general, in the non-collinear configuration (but still for $\phi \ll 1$), the Schmidt number achieves its minimal value (\ref{Kmin}), which is larger than unity, at the finite crystal length (\ref{lopt}) and under the optimal focusing of the signal beam (\ref{wopt}). Both optimal values are proportional to the spatial length of the gate pulse in the crystal $\t_g/k_s'$.
In turn, in a slightly non-collinear configuration defined by the condition
	\begin{align}
	& \phi^2 + |\phi (\phi-\rho)| \ll \phi_0^2 \L{f0_cond}
	\end{align}
and for the optimally focused signal beam, the dependence of the Schmidt number (\ref{Klw}) on the crystal length (when $l<l_{\text{opt}}$) takes the approximate form
	\begin{align}
	& K(l,w_s) \big\vert_{w_s=w_{opt}} = \sqrt{1+\({l_0}/{l}\)^2}
	\end{align}
where we defined characteristic length of the temporal walk-off of signal and up-converted pulses in the crystal
	\begin{align}
	& l_0 = \frac{\t_g}{\sqrt{\g/2} \; (k_c'-k_s')} \L{l0}
	\end{align}
In this case a condition for the single-mode regime reads
	\begin{align}
	& l \gg l_0 \L{l0_cond}
	\end{align}
The condition requires strong walk-off of signal and up-converted pulses along the crystal length. For given pulses, it is achieved by choosing longer crystals with a larger difference of group velocities for the fundamental and doubled frequencies. It is similar to the long crystal condition for the generation of the decorrelated photon pair in a parametric down-conversion scheme \cite{U'Ren2006}.  At the same time, for longer crystals the effects of diffraction and higher order dispersion may become important.~\footnote{For  an optical pulse at 800 nm with the spectral width $\Delta \l = 6$ nm (FWHM) and waist radius $0.1$ mm, the Rayleigh length is $l_R \sim 40$ mm and the dispersion length $l_D \sim 60$ mm in a BBO crystal.}

Under the conditions (\ref{f0_cond}) and (\ref{l0_cond}) the transfer function (\ref{LWqW}) is factorized
	\begin{align}
	& L(\W_c, q_c, \W_s) = \l \; \psi(\W_c, q_c) \; \varphi(\W_s)
	\end{align}
giving the spectral profiles of the subtracted $\varphi$ and up-converted $\psi$ modes
	\begin{align}
	& \varphi(\W_s) = \a_g(\W_s),\\
	& \psi(\W_c,q_c) \propto u_s(q_c) \; \sinc\bigg\{\Big[(k'_c-k'_s) \W_c+(\phi-\rho)q_c\Big] l/2\bigg\}
	\end{align}
One sees that the profile of the subtracted mode is defined by the gating pulse, thus providing a very convenient control over the photon subtraction mode. In turn, the up-converted spatio-temporal mode is defined by the spatial profile of the optimally focused signal beam and the phase matching function. It reveals angular dispersion, i.e. correlations between frequency and transverse momentum: $(\rho-\phi) q_c \approx (k'_c-k'_s) \W_c$. These correlations become even stronger approaching the single-mode regime (increasing crystal length).
The only Schmidt coefficient in the single-mode regime reads
	\begin{align}
	\l = \sqrt{\frac{\pi}{(k_c'-k_s') \, l/2}} \L{l}
	\end{align}
Let us estimate the probability of photon subtraction in the single-mode regime when the only subtraction mode is matched to the signal mode of the incoming field. The expression (\ref{P_multi}) gives the probability $P_{1} = |C'|^2 \l^2 N_s$ which is proportional to the up-conversion efficiency of the mode and number of photons in this mode $N_s$. For a Gaussian distribution of the transverse profile of gate (\ref{u}), the proportionality coefficient reads $C' = 2\sqrt{\pi} \; C/w_g$. Using (\ref{l}), one gets the explicit expression for the probability that we normalize to the number of photons in the mode and average energy of the gate pulse per unit area
	\begin{align}
	& \frac{P_{1,\text{Gauss}}}{N_s \; W_g/\pi w_g^2} = \frac{\(\varepsilon_0 \chi^{(2)} {\cal E}_s {\cal E}_c\)^2}{ 2\varepsilon_0 n_g c \; \hbar^2}  \frac{l}{k_c'-k_s'} \L{Nsqz}
	\end{align}

Using analytical results and the experimental parameters depicted in Appendix \ref{Exp}, one gets the following estimations for the single mode regime: the walk-off length $l_0=1.6$ mm; non-collinear angle $\phi_0 = 8^\circ $. The normalized probability of the photon subtraction is $\sim 0.2$ per J/mm$^2$.
We compare these values with the numerical results of the following section.

\section{Modeling photon subtraction from a squeezed frequency comb}\L{Num}

In this section, we calculate numerically the number of subtraction modes and their profiles for realistic parameters beyond the Gaussian approximation of the transfer function (\ref{LWqW}). This is performed by the eigen decomposition of the function $G(\W_s,\W_s') \equiv \int \ud\W_c \ud q_c \; L(\W_c, q_c, \W_s) \; L^*(\W_c, q_c, \W'_s)$ introduced in Eq. (\ref{G_def}). In that case, the eigen-functions correspond to the subtraction modes and the eigen-values are equal to the squared Schmidt coefficients.
 In order to model photon subtraction from a multimode squeezed frequency comb generated within an optical cavity, we choose the experimental parameters of \cite{Roslund2013}. See Appendix \ref{Exp} for more details on these parameters.

Fig. \ref{fig:K=K(w,l,G)} represents the number of subtraction modes (the Schmidt number) for Gaussian gate pulses. It corresponds to the analytical case studied in the previous section. The number of modes depends on the crystal length, the focus of the signal beam (i.e. waist radius) and the non-collinear angle. The two graphs on the left represent the case where the signal pulse propagates in the same direction as the up-converted pulse (i.e. $\phi$ and $\rho$ have the same sign) and the graph on the right is for the opposite configuration. The dashed vertical line marks the characteristic length of the temporal walk-off of pulses (\ref{l0}).
For small non-collinear angles $\phi \ll \phi_0 $ (top plots) and a crystal length larger then the temporal walk-off length, the Schmidt number tends to unity. For larger non-collinear angles $\phi \sim \phi_0$ (bottom plots), a minimal Schmidt value is higher than unity and is achieved for a crystal length close to the walk-off length.
This behavior is in good agreement with the analytical prediction.
Also analytical expressions give good quantitative estimations for a minimum of the Schmidt number when  $\phi$ and $\rho$ have opposite signs (right plots), as it is depicted in the figure caption.
In both cases there is an optimal focusing of the signal beam that delivers minimal Schmidt number for the given crystal length. This focusing is well approximated by its analytical value (\ref{wopt}). It is indicated by the dashed horizontal line on the figures.
One sees that switching to an even more non-collinear configuration demands a more careful focusing of the signal beam in order to stay at the minimum of the Schmidt number. Furthermore, when the value of the non-collinear angle is close to the walk-off angle, then the difference between the two configurations (co-propagation of the pulses and opposite propagation) becomes significant.

Fig. \ref{fig:K=K(w,l,HG1)} represents the number of subtraction modes for different spectral shapes of the gate pulse, namely zero, first and second order Hermite-Gaussian modes. The crystal length is set at 2 mm. One sees that the dependence on signal beam focusing is preserved for higher order transverse profiles and that the minimum location is still well approximated by the analytical expression. However, the multimode nature of the up-conversion process increases significantly with the concomitant increase of the gate order. It means in particular that analytical single-mode conditions (\ref{f0_cond}) and (\ref{l0_cond}) have to be adjusted for higher order gates.

\begin{figure}
	\center{\includegraphics[width=0.95\linewidth]{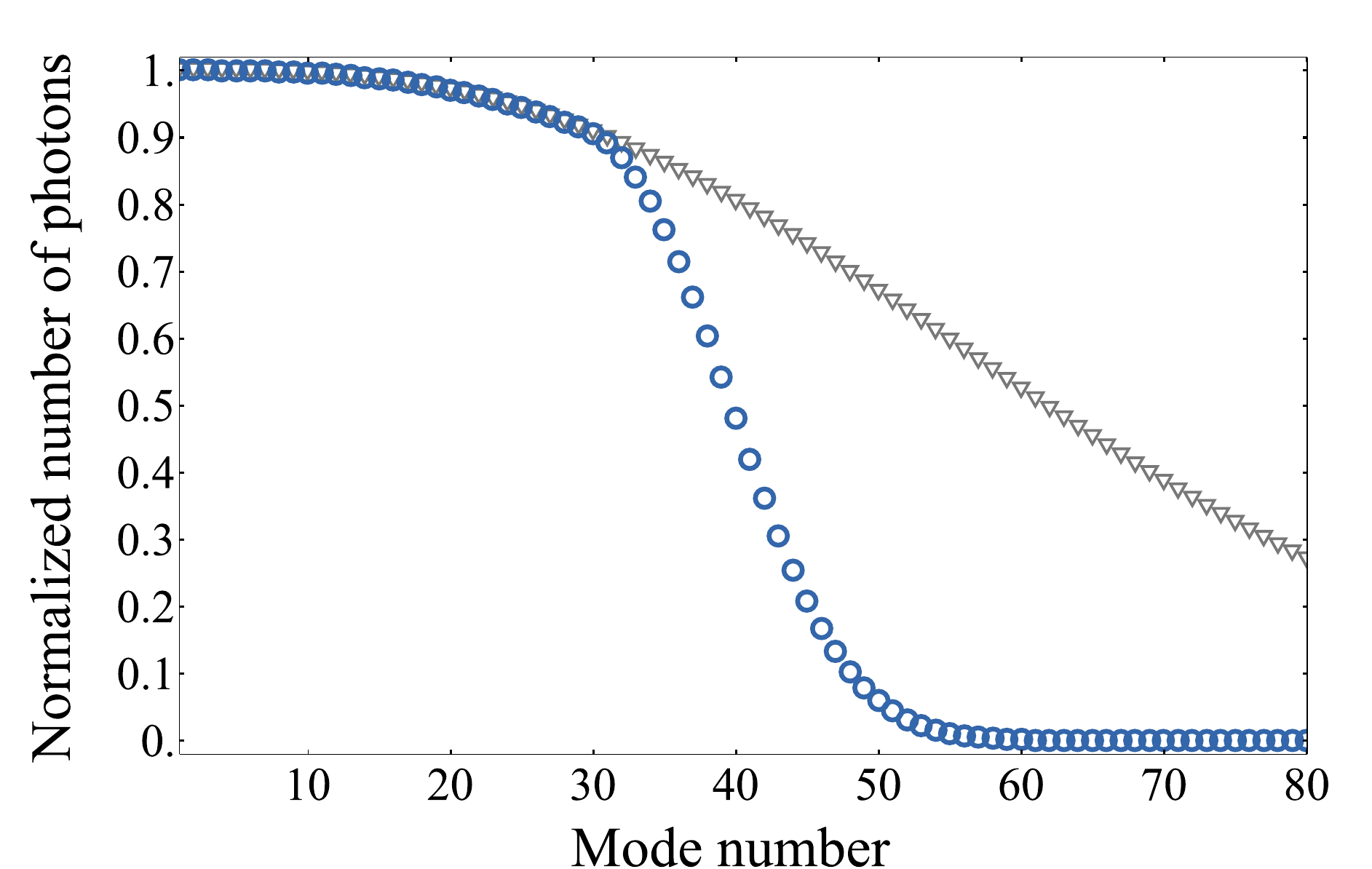}}
	\caption{Normalized average number of photons in eigen-modes of the squeezed frequency comb estimated for the experimental parameters \cite{Roslund2013}: without filtering in the optical parametric oscillator (triangles); with 100 nm bandwidth filtering (circles).
Number of photons in single pulses of modes can be estimated multiplying the distribution with $ 6\times 10^{-3}$ photons that corresponds to the measured 4.2 dB squeezing of the first mode and the cavity finesse which is equal to ${\cal F}=40$.}
	\label{fig:Ns}
	\end{figure}

\begin{figure*}
	\begin{minipage}[h]{1.\linewidth}
	\center{\includegraphics[width=1\linewidth]{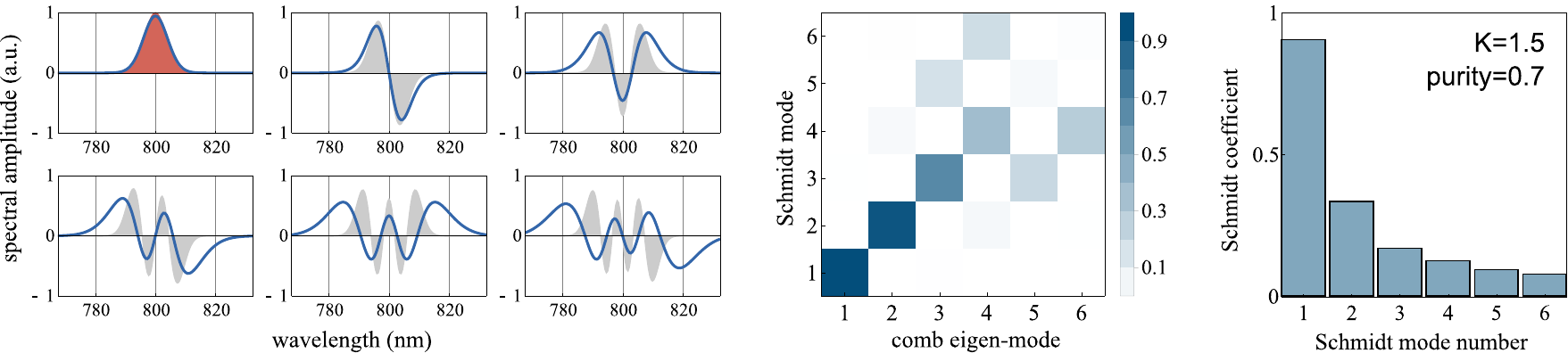} \\ a) gate pulse = HG-0 }
	\end{minipage}
	\hspace{1 cm}
	\begin{minipage}[h]{1.\linewidth}
	\center{\includegraphics[width=1\linewidth]{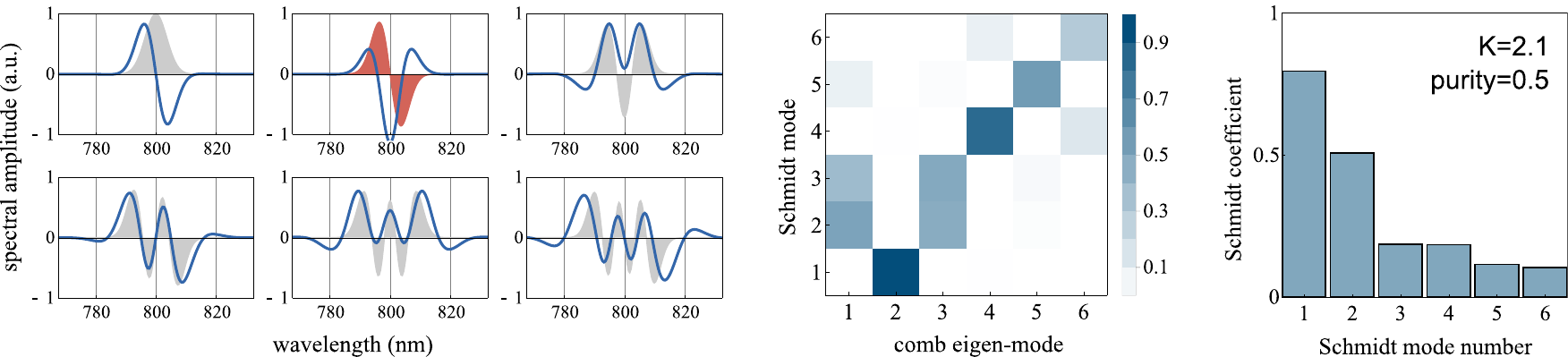} \\ b) gate pulse = HG-1 }
	\end{minipage}
	\vfill
	\begin{minipage}[h]{1.\linewidth}
	\center{\includegraphics[width=1\linewidth]{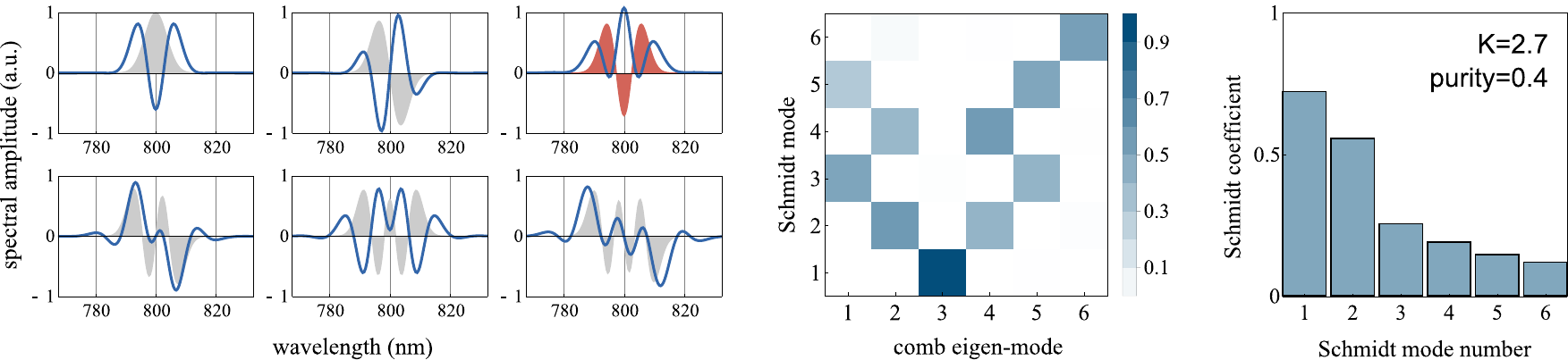} \\ c) gate pulse = HG-2 }\\
	\end{minipage}
	\caption{Photon subtraction from the multimode squeezed frequency comb via the parametric up-conversion process.
From left to right: first six subtraction modes (solid blue lines) in comparison with  the Hermite-Gaussian eigen-modes of the comb (filled grey curves); matrix of overlap coefficients; distribution of Schmidt coefficients. 
From top to bottom: gate pulse (filled red/dark grey curves) is matched successively to the first three eigen-modes of the comb.
Insets indicate Schmidt number, purity of the photon subtracted state and photon subtraction rate. Parameters of the non-collinear arrangement are marked in Fig. \ref{fig:K=K(w,l,HG1)}~(b).}
	\label{fig:state}
	\end{figure*}

These results indicate how to optimally choose the parameters to approach the single-mode regime in the photon-subtraction arrangement. This regime is sufficient for a pure conditioned state of the signal field after a photon subtraction.
Let us now model photon subtraction from a highly multimode squeezed frequency comb.
Fig. \ref{fig:Ns} represents distribution of photons per eigen-modes of the comb estimated for experimental conditions. Details of this estimation are presented in the Appendix \ref{Exp}.
In the temporal domain, the comb corresponds to a train of correlated pulses \cite{Averchenko2010, Jiang2012c} where the number of pulses is roughly equal to the finesse of the optical cavity.
We consider the subtraction of a photon from a single pulse of the train.
We approximate the eigen-modes of signal by Hermite-Gaussians of different orders $\{s_n(\w_s)\} = \{\text{HG}_n(\w_s)\}$ with a characteristic bandwidth of 6 nm. Then the spectral profile of the gate pulse is consequently set as the first three HG modes, thus matching the subtraction mode with the target mode from the comb.
The corresponding working points of the up-conversion for each gate profile are marked in Fig.~\ref{fig:K=K(w,l,HG1)}~(b).
Then Fig.~\ref{fig:state} represents the first six eigen-modes of the comb (gray filled curves), subtraction modes (blue solid lines) for different gate pulses (red filled curves) and overlap matrices $|O_{mn}|^2=\left|\int \varphi^*_m(\w_s) s_n(\w_s) \ud\w_s\right|^2$ between these modes. One sees that the first subtraction mode is defined by the gate pulse shape.
The right column of the figure represents the Schmidt coefficients distribution, i.e. the modes that are involved into the up-conversion.
Since the up-conversion is not a perfectly single-mode process for the chosen parameters, the purity of the conditioned state is consequently degraded.
The Schmidt number and the purity of the output state calculated with expression (\ref{purity}) are indicated in the figure as inset.
While the purity successively decreases for higher order extraction modes the rate of the photon subtraction (calculated from the expression (\ref{P_multi})) does not change and equals to 370 s$^{-1}$ for the following parameters: gate pulse energy $W_g = 10$ nJ, repetition rate of pulses 80 MHz; gate waist radius $w_g = 1\mbox{ mm} \gg w_s=0.1\mbox{ mm}$ (fulfilling the plane wave approximation). The value coincides with the analytical estimation (\ref{Nsqz}) for the same parameters. This stability of the photon subtraction rate is a consequence of almost flat distribution of photons in eigen-modes of the comb (see Fig.~\ref{fig:Ns}).

\begin{figure*}
	\center{\includegraphics[width=1\linewidth]{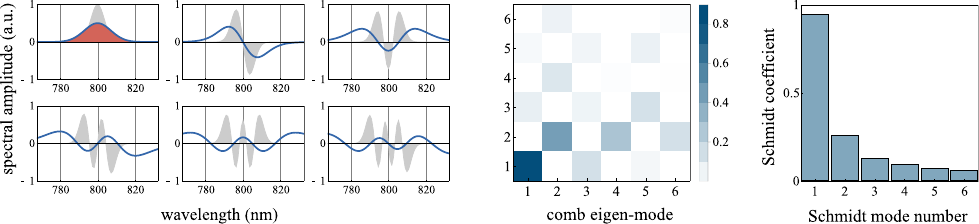}}
	\caption{The same as Fig. \ref{fig:state}-a but for a Gaussian gate pulse twice spectrally broader.}
	\label{fig:state-1}
	\end{figure*}

It was pointed out that the conditioned state of the signal field depends on the matching between the eigen-modes of the field and the subtraction modes. Fig. \ref{fig:state-1} illustrates the situation when the subtraction modes are two times spectrally broader than above. It is achieved by using broader Gaussian gate pulses. It leads to a more selective overlap matrix. Furthermore, the up-conversion becomes closer to the single-mode regime, as it seen from the value of the Schmidt number. As a result, state purity is increased.

\section{Conclusion}\L{Concl}

In this article, we have shown that it is possible to use non-collinear sum-frequency generation to achieve a mode-dependent beamsplitter in the continuous variable regime and perform mode selective photon subtraction. Using both analytical and numerical analysis, we have demonstrated under which condition one can minimize the Schmidt number of the process. The mode from which the photon is subtracted can be controlled via proper pump pulse shaping, even though Schmidt number increases with higher order Hermite modes. 

Recently mode selective parametric up-conversion has been demonstrated in a non-linear optical waveguide \cite{Brecht2014}. Obtained results confirms experimental feasibility of the proposed photon subtraction method.  

Proposed method, applied to multimode frequency comb, could be applied to any broadband source of quantum states. Furthermore, taking into account spatio-temporal distribution of an up-converted photon, one can apply additional filtering to further improve and control the Schmidt number of the process.

\acknowledgments
We thank Christine Silberhorn and Benjamin Brecht for useful discussions. V.A. acknowledges support of Ville de Paris. This work is supported by the European Research Council starting grant Frecquam.

\begin{figure*}[H]
\begin{minipage}[h]{\length\linewidth}
\center{\includegraphics[width=0.8\linewidth]{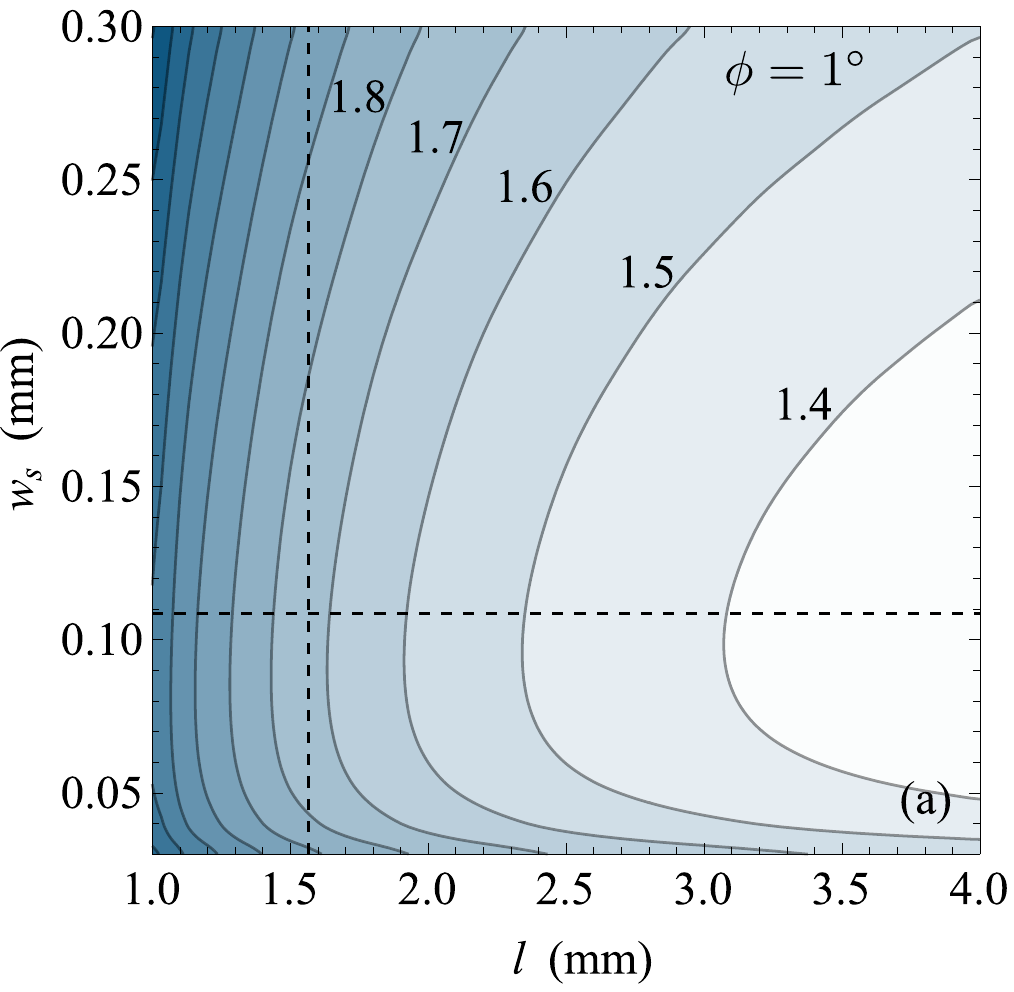}} \\
\end{minipage}
\hfill
\begin{minipage}[h]{\length\linewidth}
\center{\includegraphics[width=0.8\linewidth]{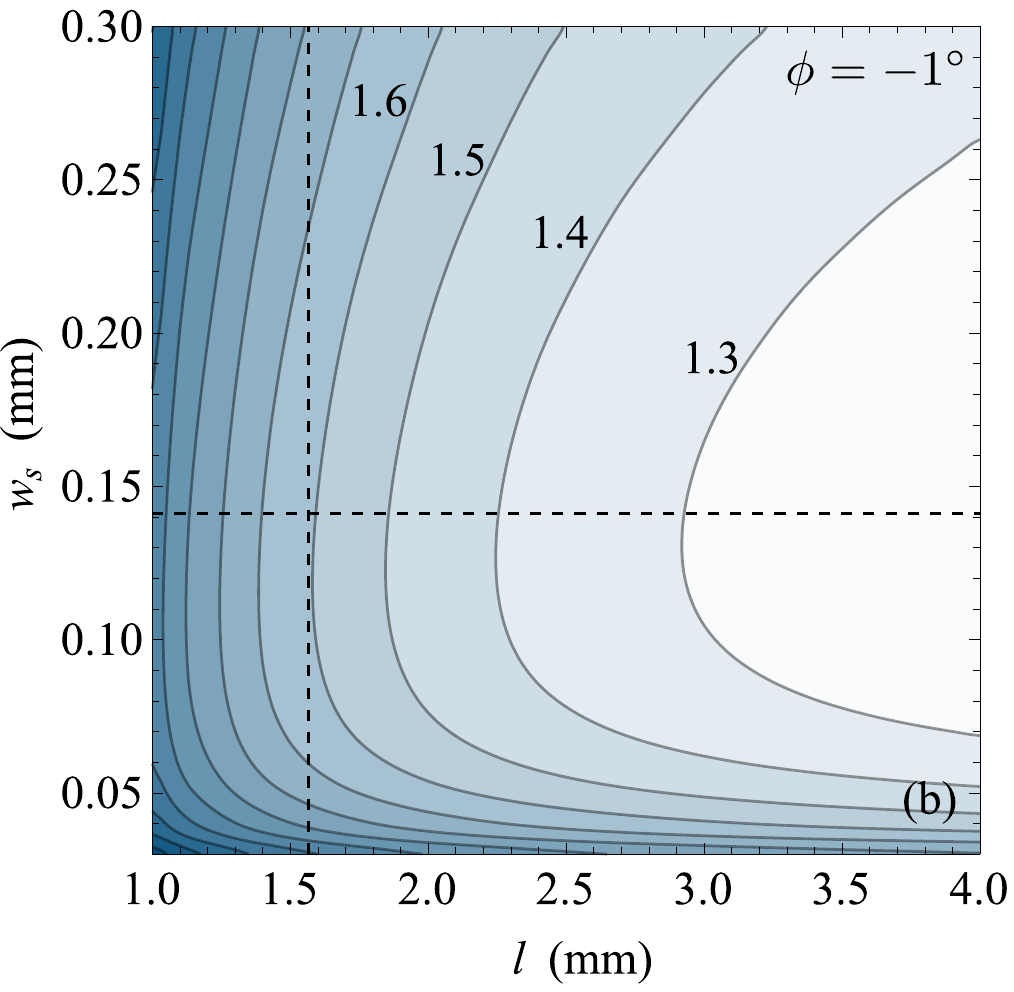}} \\
\end{minipage}
\vfill
\begin{minipage}[h]{\length\linewidth}
\center{\includegraphics[width=0.8\linewidth]{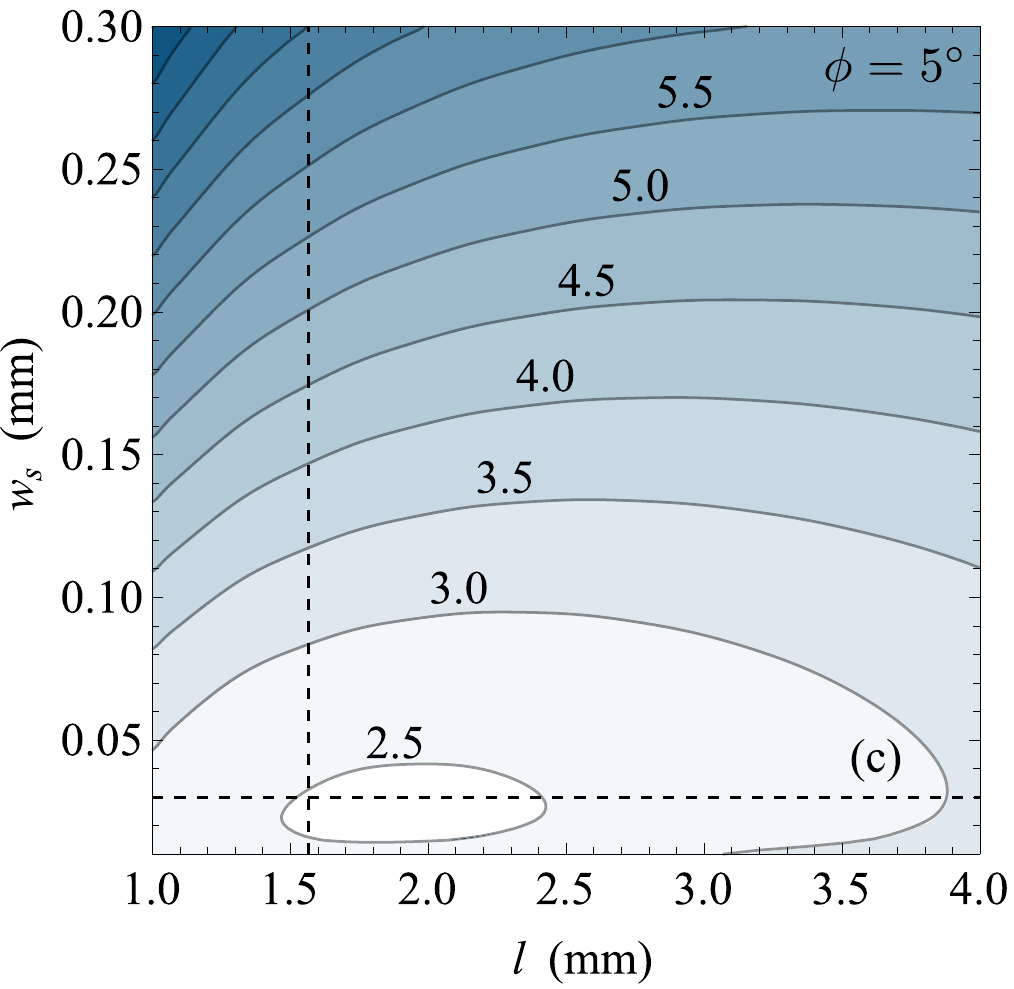}} \\
\end{minipage}
\hfill
\begin{minipage}[h]{\length\linewidth}
\center{\includegraphics[width=0.8\linewidth]{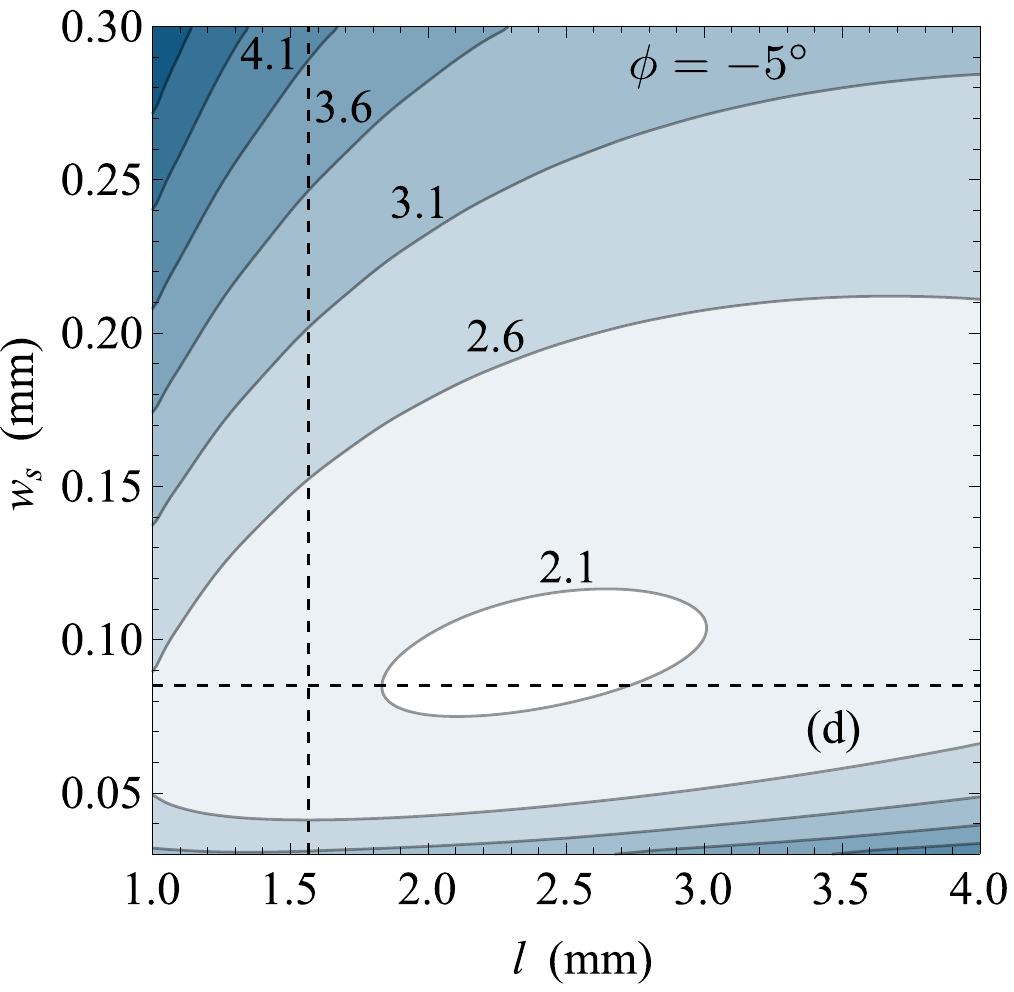}}
\end{minipage}
\caption{Schmidt number $K$ as a function of the crystal length $l$ and signal beam focusing $w_s$ for different non-collinear angles $\phi$. These plots are obtained for a Gaussian spectral profile of the gate pulses. Dashed horizontal line indicates the analytical estimation of the optimal focusing $w_{\text{opt}}$ (see (\ref{wopt})). Vertical line depicts the length $l_0$ of temporal walk-off of pulses (see (\ref{l0})). Characteristic non-collinear angle (\ref{f0}) is $\phi_0 = 8^\circ$. Analytical estimations for $K_{\text{min}}$ (\ref{Kmin}) and $l_{\text{opt}}$ (\ref{lopt}), respectively: 1.1 at 11 mm  (a, b); 1.5 at 2.4 mm (c);  2.2 at 2.4 mm (d).}
\label{fig:K=K(w,l,G)}
\end{figure*}
\begin{figure*}[H]
\begin{minipage}[h]{\length\linewidth}
\center{\includegraphics[width=0.8\linewidth]{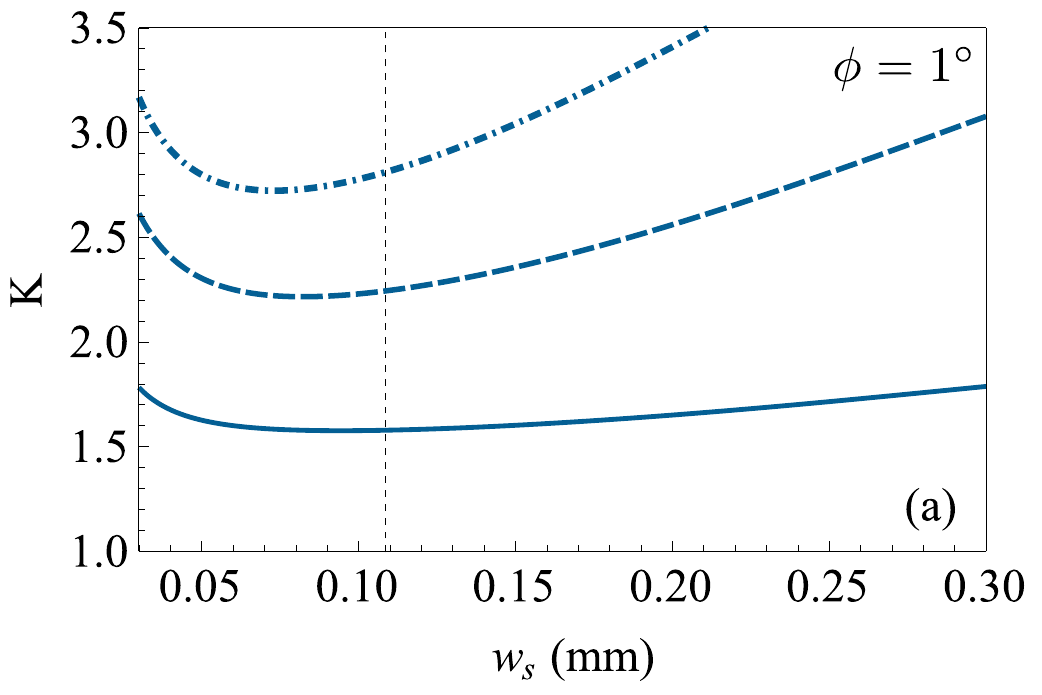}} \\ 
\end{minipage}
\hfill
\begin{minipage}[h]{\length\linewidth}
\center{\includegraphics[width=0.8\linewidth]{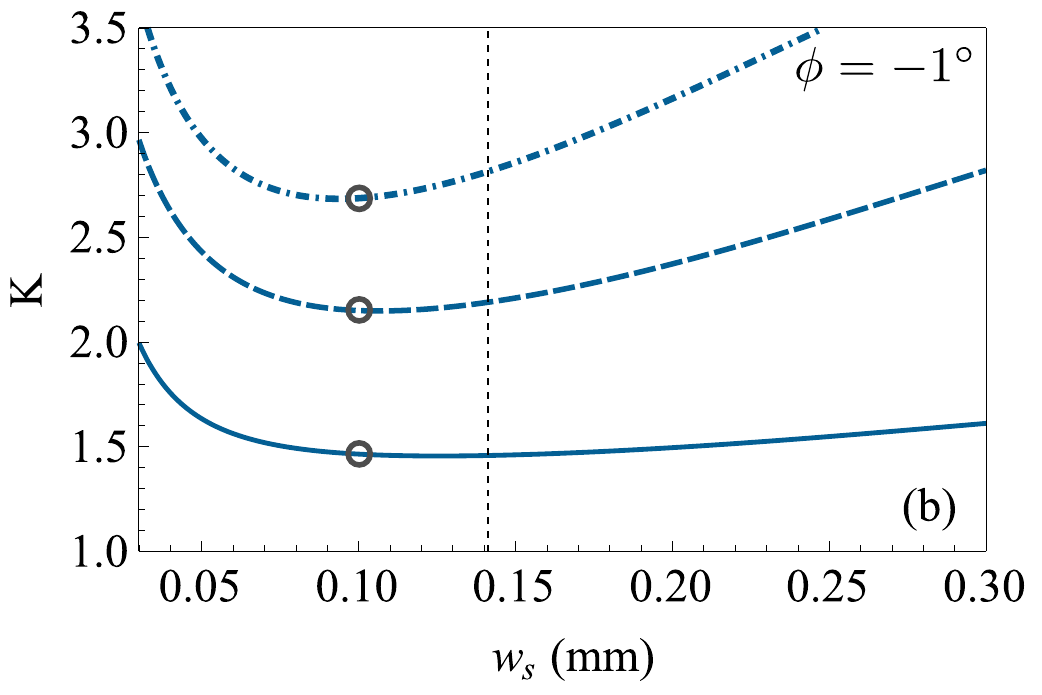}} \\
\end{minipage}
\vfill
\begin{minipage}[h]{\length\linewidth}
\center{\includegraphics[width=0.8\linewidth]{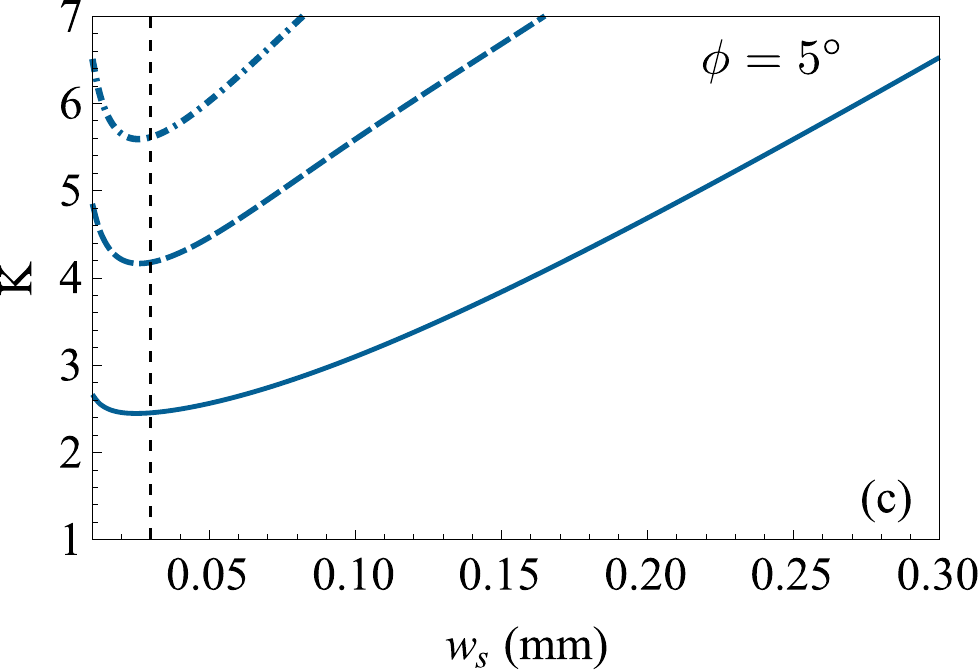}} \\
\end{minipage}
\hfill
\begin{minipage}[h]{\length\linewidth}
\center{\includegraphics[width=0.8\linewidth]{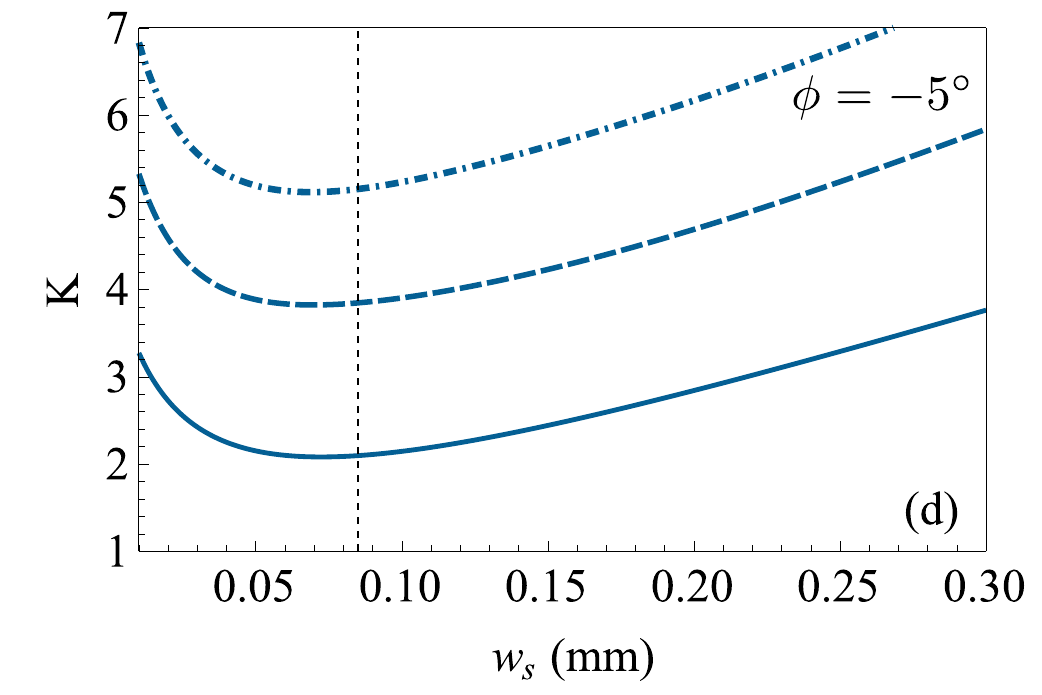}}
\end{minipage}
\caption{Schmidt number $K$ as a function of the signal beam focusing $w_s$  for different non-collinear angles and Hermite-Gaussian gating pulses of different orders: HG-0 (solid line), HG-1 (dashed line) and HG-2 (dashed-dotted line). BBO crystal length is $l=2$ mm. Dashed vertical line indicates analytically estimated optimal focusing $w_{\text{opt}}$ in the long crystal limit. Circles at the figure (b) depict working points of the up-conversion process chosen to model a photon subtraction from the squeezed frequency comb.}
\label{fig:K=K(w,l,HG1)}
\end{figure*}

\bibliography{NonlinearPhotonSubtractionBibl}

\begin{thebibliography}{35}%
\makeatletter
\providecommand \@ifxundefined [1]{%
 \@ifx{#1\undefined}
}%
\providecommand \@ifnum [1]{%
 \ifnum #1\expandafter \@firstoftwo
 \else \expandafter \@secondoftwo
 \fi
}%
\providecommand \@ifx [1]{%
 \ifx #1\expandafter \@firstoftwo
 \else \expandafter \@secondoftwo
 \fi
}%
\providecommand \natexlab [1]{#1}%
\providecommand \enquote  [1]{``#1''}%
\providecommand \bibnamefont  [1]{#1}%
\providecommand \bibfnamefont [1]{#1}%
\providecommand \citenamefont [1]{#1}%
\providecommand \href@noop [0]{\@secondoftwo}%
\providecommand \href [0]{\begingroup \@sanitize@url \@href}%
\providecommand \@href[1]{\@@startlink{#1}\@@href}%
\providecommand \@@href[1]{\endgroup#1\@@endlink}%
\providecommand \@sanitize@url [0]{\catcode `\\12\catcode `\$12\catcode
  `\&12\catcode `\#12\catcode `\^12\catcode `\_12\catcode `\%12\relax}%
\providecommand \@@startlink[1]{}%
\providecommand \@@endlink[0]{}%
\providecommand \url  [0]{\begingroup\@sanitize@url \@url }%
\providecommand \@url [1]{\endgroup\@href {#1}{\urlprefix }}%
\providecommand \urlprefix  [0]{URL }%
\providecommand \Eprint [0]{\href }%
\providecommand \doibase [0]{http://dx.doi.org/}%
\providecommand \selectlanguage [0]{\@gobble}%
\providecommand \bibinfo  [0]{\@secondoftwo}%
\providecommand \bibfield  [0]{\@secondoftwo}%
\providecommand \translation [1]{[#1]}%
\providecommand \BibitemOpen [0]{}%
\providecommand \bibitemStop [0]{}%
\providecommand \bibitemNoStop [0]{.\EOS\space}%
\providecommand \EOS [0]{\spacefactor3000\relax}%
\providecommand \BibitemShut  [1]{\csname bibitem#1\endcsname}%
\let\auto@bib@innerbib\@empty
\bibitem [{\citenamefont {Lewenstein}\ \emph {et~al.}(2007)\citenamefont
  {Lewenstein}, \citenamefont {Sanpera}, \citenamefont {Ahufinger},
  \citenamefont {Damski}, \citenamefont {Sen(De)},\ and\ \citenamefont
  {Sen}}]{Lewenstein2007}%
  \BibitemOpen
  \bibfield  {author} {\bibinfo {author} {\bibfnamefont {M.}~\bibnamefont
  {Lewenstein}}, \bibinfo {author} {\bibfnamefont {A.}~\bibnamefont {Sanpera}},
  \bibinfo {author} {\bibfnamefont {V.}~\bibnamefont {Ahufinger}}, \bibinfo
  {author} {\bibfnamefont {B.}~\bibnamefont {Damski}}, \bibinfo {author}
  {\bibfnamefont {A.}~\bibnamefont {Sen(De)}}, \ and\ \bibinfo {author}
  {\bibfnamefont {U.}~\bibnamefont {Sen}},\ }\href {\doibase
  10.1080/00018730701223200} {\bibfield  {journal} {\bibinfo  {journal}
  {Advances in Physics}\ }\textbf {\bibinfo {volume} {56}},\ \bibinfo {pages}
  {243} (\bibinfo {year} {2007})}\BibitemShut {NoStop}%
\bibitem [{\citenamefont {Haffner}\ \emph {et~al.}(2008)\citenamefont
  {Haffner}, \citenamefont {Roos},\ and\ \citenamefont {Blatt}}]{HAFFNER2008}%
  \BibitemOpen
  \bibfield  {author} {\bibinfo {author} {\bibfnamefont {H.}~\bibnamefont
  {Haffner}}, \bibinfo {author} {\bibfnamefont {C.}~\bibnamefont {Roos}}, \
  and\ \bibinfo {author} {\bibfnamefont {R.}~\bibnamefont {Blatt}},\ }\href
  {\doibase 10.1016/j.physrep.2008.09.003} {\bibfield  {journal} {\bibinfo
  {journal} {Physics Reports}\ }\textbf {\bibinfo {volume} {469}},\ \bibinfo
  {pages} {155} (\bibinfo {year} {2008})}\BibitemShut {NoStop}%
\bibitem [{\citenamefont {Bloch}\ \emph {et~al.}(2012)\citenamefont {Bloch},
  \citenamefont {Dalibard},\ and\ \citenamefont {Nascimb\`{e}ne}}]{Bloch2012}%
  \BibitemOpen
  \bibfield  {author} {\bibinfo {author} {\bibfnamefont {I.}~\bibnamefont
  {Bloch}}, \bibinfo {author} {\bibfnamefont {J.}~\bibnamefont {Dalibard}}, \
  and\ \bibinfo {author} {\bibfnamefont {S.}~\bibnamefont {Nascimb\`{e}ne}},\
  }\href {\doibase 10.1038/nphys2259} {\bibfield  {journal} {\bibinfo
  {journal} {Nature Physics}\ }\textbf {\bibinfo {volume} {8}},\ \bibinfo
  {pages} {267} (\bibinfo {year} {2012})}\BibitemShut {NoStop}%
\bibitem [{\citenamefont {Fowler}\ \emph {et~al.}(2012)\citenamefont {Fowler},
  \citenamefont {Mariantoni}, \citenamefont {Martinis},\ and\ \citenamefont
  {Cleland}}]{Fowler2012a}%
  \BibitemOpen
  \bibfield  {author} {\bibinfo {author} {\bibfnamefont {A.~G.}\ \bibnamefont
  {Fowler}}, \bibinfo {author} {\bibfnamefont {M.}~\bibnamefont {Mariantoni}},
  \bibinfo {author} {\bibfnamefont {J.~M.}\ \bibnamefont {Martinis}}, \ and\
  \bibinfo {author} {\bibfnamefont {A.~N.}\ \bibnamefont {Cleland}},\ }\href
  {http://link.aps.org/doi/10.1103/PhysRevA.86.032324} {\bibfield  {journal}
  {\bibinfo  {journal} {Physical Review A}\ }\textbf {\bibinfo {volume} {86}},\
  \bibinfo {pages} {032324} (\bibinfo {year} {2012})}\BibitemShut {NoStop}%
\bibitem [{\citenamefont {Barends}\ \emph {et~al.}(2014)\citenamefont
  {Barends}, \citenamefont {Kelly}, \citenamefont {Megrant}, \citenamefont
  {Veitia}, \citenamefont {Sank}, \citenamefont {Jeffrey}, \citenamefont
  {White}, \citenamefont {Mutus}, \citenamefont {Fowler}, \citenamefont
  {Campbell}, \citenamefont {Chen}, \citenamefont {Chen}, \citenamefont
  {Chiaro}, \citenamefont {Dunsworth}, \citenamefont {Neill}, \citenamefont
  {O`Malley}, \citenamefont {Roushan}, \citenamefont {Vainsencher},
  \citenamefont {Wenner}, \citenamefont {Korotkov}, \citenamefont {Cleland},\
  and\ \citenamefont {Martinis}}]{Barends2014}%
  \BibitemOpen
  \bibfield  {author} {\bibinfo {author} {\bibfnamefont {R.}~\bibnamefont
  {Barends}}, \bibinfo {author} {\bibfnamefont {J.}~\bibnamefont {Kelly}},
  \bibinfo {author} {\bibfnamefont {A.}~\bibnamefont {Megrant}}, \bibinfo
  {author} {\bibfnamefont {A.}~\bibnamefont {Veitia}}, \bibinfo {author}
  {\bibfnamefont {D.}~\bibnamefont {Sank}}, \bibinfo {author} {\bibfnamefont
  {E.}~\bibnamefont {Jeffrey}}, \bibinfo {author} {\bibfnamefont {T.~C.}\
  \bibnamefont {White}}, \bibinfo {author} {\bibfnamefont {J.}~\bibnamefont
  {Mutus}}, \bibinfo {author} {\bibfnamefont {A.~G.}\ \bibnamefont {Fowler}},
  \bibinfo {author} {\bibfnamefont {B.}~\bibnamefont {Campbell}}, \bibinfo
  {author} {\bibfnamefont {Y.}~\bibnamefont {Chen}}, \bibinfo {author}
  {\bibfnamefont {Z.}~\bibnamefont {Chen}}, \bibinfo {author} {\bibfnamefont
  {B.}~\bibnamefont {Chiaro}}, \bibinfo {author} {\bibfnamefont
  {A.}~\bibnamefont {Dunsworth}}, \bibinfo {author} {\bibfnamefont
  {C.}~\bibnamefont {Neill}}, \bibinfo {author} {\bibfnamefont
  {P.}~\bibnamefont {O`Malley}}, \bibinfo {author} {\bibfnamefont
  {P.}~\bibnamefont {Roushan}}, \bibinfo {author} {\bibfnamefont
  {A.}~\bibnamefont {Vainsencher}}, \bibinfo {author} {\bibfnamefont
  {J.}~\bibnamefont {Wenner}}, \bibinfo {author} {\bibfnamefont {A.~N.}\
  \bibnamefont {Korotkov}}, \bibinfo {author} {\bibfnamefont {A.~N.}\
  \bibnamefont {Cleland}}, \ and\ \bibinfo {author} {\bibfnamefont {J.~M.}\
  \bibnamefont {Martinis}},\ }\href {http://arxiv.org/abs/1402.4848} {\bibfield
   {journal} {\bibinfo  {journal} {arXiv:1402.4848}\ } (\bibinfo {year}
  {2014})}\BibitemShut {NoStop}%
\bibitem [{\citenamefont {Walther}\ \emph {et~al.}(2005)\citenamefont
  {Walther}, \citenamefont {Resch}, \citenamefont {Rudolph}, \citenamefont
  {Schenck}, \citenamefont {Weinfurter}, \citenamefont {Vedral}, \citenamefont
  {Aspelmeyer},\ and\ \citenamefont {Zeilinger}}]{Walther2005}%
  \BibitemOpen
  \bibfield  {author} {\bibinfo {author} {\bibfnamefont {P.}~\bibnamefont
  {Walther}}, \bibinfo {author} {\bibfnamefont {K.~J.}\ \bibnamefont {Resch}},
  \bibinfo {author} {\bibfnamefont {T.}~\bibnamefont {Rudolph}}, \bibinfo
  {author} {\bibfnamefont {E.}~\bibnamefont {Schenck}}, \bibinfo {author}
  {\bibfnamefont {H.}~\bibnamefont {Weinfurter}}, \bibinfo {author}
  {\bibfnamefont {V.}~\bibnamefont {Vedral}}, \bibinfo {author} {\bibfnamefont
  {M.}~\bibnamefont {Aspelmeyer}}, \ and\ \bibinfo {author} {\bibfnamefont
  {A.}~\bibnamefont {Zeilinger}},\ }\href {\doibase 10.1038/nature03347}
  {\bibfield  {journal} {\bibinfo  {journal} {Nature}\ }\textbf {\bibinfo
  {volume} {434}},\ \bibinfo {pages} {169} (\bibinfo {year}
  {2005})}\BibitemShut {NoStop}%
\bibitem [{\citenamefont {Aoki}\ \emph {et~al.}(2009)\citenamefont {Aoki},
  \citenamefont {Takahashi}, \citenamefont {Kajiya}, \citenamefont {Yoshikawa},
  \citenamefont {Braunstein}, \citenamefont {van Loock},\ and\ \citenamefont
  {Furusawa}}]{Aoki2009}%
  \BibitemOpen
  \bibfield  {author} {\bibinfo {author} {\bibfnamefont {T.}~\bibnamefont
  {Aoki}}, \bibinfo {author} {\bibfnamefont {G.}~\bibnamefont {Takahashi}},
  \bibinfo {author} {\bibfnamefont {T.}~\bibnamefont {Kajiya}}, \bibinfo
  {author} {\bibfnamefont {J.}~\bibnamefont {Yoshikawa}}, \bibinfo {author}
  {\bibfnamefont {S.~L.}\ \bibnamefont {Braunstein}}, \bibinfo {author}
  {\bibfnamefont {P.}~\bibnamefont {van Loock}}, \ and\ \bibinfo {author}
  {\bibfnamefont {A.}~\bibnamefont {Furusawa}},\ }\href {\doibase
  10.1038/nphys1309} {\bibfield  {journal} {\bibinfo  {journal} {Nature
  Physics}\ }\textbf {\bibinfo {volume} {5}},\ \bibinfo {pages} {541} (\bibinfo
  {year} {2009})}\BibitemShut {NoStop}%
\bibitem [{\citenamefont {Ukai}\ \emph {et~al.}(2011)\citenamefont {Ukai},
  \citenamefont {Iwata}, \citenamefont {Shimokawa}, \citenamefont {Armstrong},
  \citenamefont {Politi}, \citenamefont {Yoshikawa}, \citenamefont {van
  Loock},\ and\ \citenamefont {Furusawa}}]{Ukai2011}%
  \BibitemOpen
  \bibfield  {author} {\bibinfo {author} {\bibfnamefont {R.}~\bibnamefont
  {Ukai}}, \bibinfo {author} {\bibfnamefont {N.}~\bibnamefont {Iwata}},
  \bibinfo {author} {\bibfnamefont {Y.}~\bibnamefont {Shimokawa}}, \bibinfo
  {author} {\bibfnamefont {S.~C.}\ \bibnamefont {Armstrong}}, \bibinfo {author}
  {\bibfnamefont {A.}~\bibnamefont {Politi}}, \bibinfo {author} {\bibfnamefont
  {J.~I.}\ \bibnamefont {Yoshikawa}}, \bibinfo {author} {\bibfnamefont
  {P.}~\bibnamefont {van Loock}}, \ and\ \bibinfo {author} {\bibfnamefont
  {A.}~\bibnamefont {Furusawa}},\ }\href
  {http://link.aps.org/doi/10.1103/PhysRevLett.106.240504} {\bibfield
  {journal} {\bibinfo  {journal} {Physical Review Letters}\ }\textbf {\bibinfo
  {volume} {106}},\ \bibinfo {pages} {240504} (\bibinfo {year}
  {2011})}\BibitemShut {NoStop}%
\bibitem [{\citenamefont {Yokoyama}\ \emph {et~al.}(2013)\citenamefont
  {Yokoyama}, \citenamefont {Ukai}, \citenamefont {Armstrong}, \citenamefont
  {Sornphiphatphong}, \citenamefont {Kaji}, \citenamefont {Suzuki},
  \citenamefont {Yoshikawa}, \citenamefont {Yonezawa}, \citenamefont
  {Menicucci},\ and\ \citenamefont {Furusawa}}]{Yokoyama2013}%
  \BibitemOpen
  \bibfield  {author} {\bibinfo {author} {\bibfnamefont {S.}~\bibnamefont
  {Yokoyama}}, \bibinfo {author} {\bibfnamefont {R.}~\bibnamefont {Ukai}},
  \bibinfo {author} {\bibfnamefont {S.~C.}\ \bibnamefont {Armstrong}}, \bibinfo
  {author} {\bibfnamefont {C.}~\bibnamefont {Sornphiphatphong}}, \bibinfo
  {author} {\bibfnamefont {T.}~\bibnamefont {Kaji}}, \bibinfo {author}
  {\bibfnamefont {S.}~\bibnamefont {Suzuki}}, \bibinfo {author} {\bibfnamefont
  {J.}~\bibnamefont {Yoshikawa}}, \bibinfo {author} {\bibfnamefont
  {H.}~\bibnamefont {Yonezawa}}, \bibinfo {author} {\bibfnamefont {N.~C.}\
  \bibnamefont {Menicucci}}, \ and\ \bibinfo {author} {\bibfnamefont
  {A.}~\bibnamefont {Furusawa}},\ }\href {\doibase 10.1038/nphoton.2013.287}
  {\bibfield  {journal} {\bibinfo  {journal} {Nature Photonics}\ }\textbf
  {\bibinfo {volume} {7}},\ \bibinfo {pages} {982} (\bibinfo {year}
  {2013})}\BibitemShut {NoStop}%
\bibitem [{\citenamefont {Pysher}\ \emph {et~al.}(2011)\citenamefont {Pysher},
  \citenamefont {Miwa}, \citenamefont {Shahrokhshahi}, \citenamefont
  {Bloomer},\ and\ \citenamefont {Pfister}}]{Pysher2011}%
  \BibitemOpen
  \bibfield  {author} {\bibinfo {author} {\bibfnamefont {M.}~\bibnamefont
  {Pysher}}, \bibinfo {author} {\bibfnamefont {Y.}~\bibnamefont {Miwa}},
  \bibinfo {author} {\bibfnamefont {R.}~\bibnamefont {Shahrokhshahi}}, \bibinfo
  {author} {\bibfnamefont {R.}~\bibnamefont {Bloomer}}, \ and\ \bibinfo
  {author} {\bibfnamefont {O.}~\bibnamefont {Pfister}},\ }\href {\doibase
  10.1103/PhysRevLett.107.030505} {\bibfield  {journal} {\bibinfo  {journal}
  {Physical Review Letters}\ }\textbf {\bibinfo {volume} {107}},\ \bibinfo
  {pages} {030505} (\bibinfo {year} {2011})}\BibitemShut {NoStop}%
\bibitem [{\citenamefont {Armstrong}\ \emph {et~al.}(2012)\citenamefont
  {Armstrong}, \citenamefont {Morizur}, \citenamefont {Janousek}, \citenamefont
  {Hage}, \citenamefont {Treps}, \citenamefont {Lam},\ and\ \citenamefont
  {Bachor}}]{Armstrong2012}%
  \BibitemOpen
  \bibfield  {author} {\bibinfo {author} {\bibfnamefont {S.}~\bibnamefont
  {Armstrong}}, \bibinfo {author} {\bibfnamefont {J.-F.}\ \bibnamefont
  {Morizur}}, \bibinfo {author} {\bibfnamefont {J.}~\bibnamefont {Janousek}},
  \bibinfo {author} {\bibfnamefont {B.}~\bibnamefont {Hage}}, \bibinfo {author}
  {\bibfnamefont {N.}~\bibnamefont {Treps}}, \bibinfo {author} {\bibfnamefont
  {P.~K.}\ \bibnamefont {Lam}}, \ and\ \bibinfo {author} {\bibfnamefont
  {H.-A.}\ \bibnamefont {Bachor}},\ }\href {\doibase 10.1038/ncomms2033}
  {\bibfield  {journal} {\bibinfo  {journal} {Nature communications}\ }\textbf
  {\bibinfo {volume} {3}},\ \bibinfo {pages} {1026} (\bibinfo {year}
  {2012})}\BibitemShut {NoStop}%
\bibitem [{\citenamefont {Roslund}\ \emph {et~al.}(2014)\citenamefont
  {Roslund}, \citenamefont {de~Ara\'{u}jo}, \citenamefont {Jiang},
  \citenamefont {Fabre},\ and\ \citenamefont {Treps}}]{Roslund2013}%
  \BibitemOpen
  \bibfield  {author} {\bibinfo {author} {\bibfnamefont {J.}~\bibnamefont
  {Roslund}}, \bibinfo {author} {\bibfnamefont {R.~M.}\ \bibnamefont
  {de~Ara\'{u}jo}}, \bibinfo {author} {\bibfnamefont {S.}~\bibnamefont
  {Jiang}}, \bibinfo {author} {\bibfnamefont {C.}~\bibnamefont {Fabre}}, \ and\
  \bibinfo {author} {\bibfnamefont {N.}~\bibnamefont {Treps}},\ }\href
  {\doibase 10.1038/nphoton.2013.340} {\bibfield  {journal} {\bibinfo
  {journal} {Nature Photonics}\ }\textbf {\bibinfo {volume} {8}},\ \bibinfo
  {pages} {109} (\bibinfo {year} {2014})}\BibitemShut {NoStop}%
\bibitem [{\citenamefont {Tanzilli}\ \emph {et~al.}(2005)\citenamefont
  {Tanzilli}, \citenamefont {Tittel}, \citenamefont {Halder}, \citenamefont
  {Alibart}, \citenamefont {Baldi}, \citenamefont {Gisin},\ and\ \citenamefont
  {Zbinden}}]{Tanzilli2005}%
  \BibitemOpen
  \bibfield  {author} {\bibinfo {author} {\bibfnamefont {S.}~\bibnamefont
  {Tanzilli}}, \bibinfo {author} {\bibfnamefont {W.}~\bibnamefont {Tittel}},
  \bibinfo {author} {\bibfnamefont {M.}~\bibnamefont {Halder}}, \bibinfo
  {author} {\bibfnamefont {O.}~\bibnamefont {Alibart}}, \bibinfo {author}
  {\bibfnamefont {P.}~\bibnamefont {Baldi}}, \bibinfo {author} {\bibfnamefont
  {N.}~\bibnamefont {Gisin}}, \ and\ \bibinfo {author} {\bibfnamefont
  {H.}~\bibnamefont {Zbinden}},\ }\href {\doibase 10.1038/nature04009}
  {\bibfield  {journal} {\bibinfo  {journal} {Nature}\ }\textbf {\bibinfo
  {volume} {437}},\ \bibinfo {pages} {116} (\bibinfo {year}
  {2005})}\BibitemShut {NoStop}%
\bibitem [{\citenamefont {Vollmer}\ \emph {et~al.}(2014)\citenamefont
  {Vollmer}, \citenamefont {Baune}, \citenamefont {Samblowski}, \citenamefont
  {Eberle}, \citenamefont {H\"{a}ndchen}, \citenamefont {Fiur\'{a}\v{s}ek},\
  and\ \citenamefont {Schnabel}}]{Vollmer2014}%
  \BibitemOpen
  \bibfield  {author} {\bibinfo {author} {\bibfnamefont {C.~E.}\ \bibnamefont
  {Vollmer}}, \bibinfo {author} {\bibfnamefont {C.}~\bibnamefont {Baune}},
  \bibinfo {author} {\bibfnamefont {A.}~\bibnamefont {Samblowski}}, \bibinfo
  {author} {\bibfnamefont {T.}~\bibnamefont {Eberle}}, \bibinfo {author}
  {\bibfnamefont {V.}~\bibnamefont {H\"{a}ndchen}}, \bibinfo {author}
  {\bibfnamefont {J.}~\bibnamefont {Fiur\'{a}\v{s}ek}}, \ and\ \bibinfo
  {author} {\bibfnamefont {R.}~\bibnamefont {Schnabel}},\ }\href {\doibase
  10.1103/PhysRevLett.112.073602} {\bibfield  {journal} {\bibinfo  {journal}
  {Physical Review Letters}\ }\textbf {\bibinfo {volume} {112}},\ \bibinfo
  {pages} {073602} (\bibinfo {year} {2014})}\BibitemShut {NoStop}%
\bibitem [{\citenamefont {Kielpinski}\ \emph {et~al.}(2011)\citenamefont
  {Kielpinski}, \citenamefont {Corney},\ and\ \citenamefont
  {Wiseman}}]{Kielpinski2011}%
  \BibitemOpen
  \bibfield  {author} {\bibinfo {author} {\bibfnamefont {D.}~\bibnamefont
  {Kielpinski}}, \bibinfo {author} {\bibfnamefont {J.~F.}\ \bibnamefont
  {Corney}}, \ and\ \bibinfo {author} {\bibfnamefont {H.~M.}\ \bibnamefont
  {Wiseman}},\ }\href {\doibase 10.1103/PhysRevLett.106.130501} {\bibfield
  {journal} {\bibinfo  {journal} {Physical Review Letters}\ }\textbf {\bibinfo
  {volume} {106}},\ \bibinfo {pages} {130501} (\bibinfo {year}
  {2011})}\BibitemShut {NoStop}%
\bibitem [{\citenamefont {Lavoie}\ \emph {et~al.}(2013)\citenamefont {Lavoie},
  \citenamefont {Donohue}, \citenamefont {Wright}, \citenamefont {Fedrizzi},\
  and\ \citenamefont {Resch}}]{Lavoie2013a}%
  \BibitemOpen
  \bibfield  {author} {\bibinfo {author} {\bibfnamefont {J.}~\bibnamefont
  {Lavoie}}, \bibinfo {author} {\bibfnamefont {J.}~\bibnamefont {Donohue}},
  \bibinfo {author} {\bibfnamefont {L.}~\bibnamefont {Wright}}, \bibinfo
  {author} {\bibfnamefont {A.}~\bibnamefont {Fedrizzi}}, \ and\ \bibinfo
  {author} {\bibfnamefont {K.}~\bibnamefont {Resch}},\ }\href {\doibase
  10.1038/nphoton.2013.47} {\bibfield  {journal} {\bibinfo  {journal} {Nature
  Photonics}\ }\textbf {\bibinfo {volume} {7}},\ \bibinfo {pages} {363}
  (\bibinfo {year} {2013})}\BibitemShut {NoStop}%
\bibitem [{\citenamefont {Eckstein}\ \emph {et~al.}(2011)\citenamefont
  {Eckstein}, \citenamefont {Brecht},\ and\ \citenamefont
  {Silberhorn}}]{Eckstein2011}%
  \BibitemOpen
  \bibfield  {author} {\bibinfo {author} {\bibfnamefont {A.}~\bibnamefont
  {Eckstein}}, \bibinfo {author} {\bibfnamefont {B.}~\bibnamefont {Brecht}}, \
  and\ \bibinfo {author} {\bibfnamefont {C.}~\bibnamefont {Silberhorn}},\
  }\href {http://www.ncbi.nlm.nih.gov/pubmed/21934737} {\bibfield  {journal}
  {\bibinfo  {journal} {Optics express}\ }\textbf {\bibinfo {volume} {19}},\
  \bibinfo {pages} {13770} (\bibinfo {year} {2011})}\BibitemShut {NoStop}%
\bibitem [{\citenamefont {Huang}\ and\ \citenamefont
  {Kumar}(2013)}]{Huang2013}%
  \BibitemOpen
  \bibfield  {author} {\bibinfo {author} {\bibfnamefont {Y.-P.}\ \bibnamefont
  {Huang}}\ and\ \bibinfo {author} {\bibfnamefont {P.}~\bibnamefont {Kumar}},\
  }\href {http://www.ncbi.nlm.nih.gov/pubmed/23455105} {\bibfield  {journal}
  {\bibinfo  {journal} {Optics letters}\ }\textbf {\bibinfo {volume} {38}},\
  \bibinfo {pages} {468} (\bibinfo {year} {2013})}\BibitemShut {NoStop}%
\bibitem [{\citenamefont {Dakna}\ \emph {et~al.}(1997)\citenamefont {Dakna},
  \citenamefont {Anhut}, \citenamefont {Opatrn\'{y}}, \citenamefont
  {Kn\"{o}ll},\ and\ \citenamefont {Welsch}}]{Dakna1996}%
  \BibitemOpen
  \bibfield  {author} {\bibinfo {author} {\bibfnamefont {M.}~\bibnamefont
  {Dakna}}, \bibinfo {author} {\bibfnamefont {T.}~\bibnamefont {Anhut}},
  \bibinfo {author} {\bibfnamefont {T.}~\bibnamefont {Opatrn\'{y}}}, \bibinfo
  {author} {\bibfnamefont {L.}~\bibnamefont {Kn\"{o}ll}}, \ and\ \bibinfo
  {author} {\bibfnamefont {D.~G.}\ \bibnamefont {Welsch}},\ }\href
  {http://pra.aps.org/abstract/PRA/v55/i4/p3184\_1} {\bibfield  {journal}
  {\bibinfo  {journal} {Physical Review A}\ }\textbf {\bibinfo {volume} {55}},\
  \bibinfo {pages} {3184} (\bibinfo {year} {1997})}\BibitemShut {NoStop}%
\bibitem [{\citenamefont {Wenger}\ \emph {et~al.}(2004)\citenamefont {Wenger},
  \citenamefont {Tualle-Brouri},\ and\ \citenamefont {Grangier}}]{Wenger2004}%
  \BibitemOpen
  \bibfield  {author} {\bibinfo {author} {\bibfnamefont {J.}~\bibnamefont
  {Wenger}}, \bibinfo {author} {\bibfnamefont {R.}~\bibnamefont
  {Tualle-Brouri}}, \ and\ \bibinfo {author} {\bibfnamefont {P.}~\bibnamefont
  {Grangier}},\ }\href {\doibase 10.1103/PhysRevLett.92.153601} {\bibfield
  {journal} {\bibinfo  {journal} {Physical Review Letters}\ }\textbf {\bibinfo
  {volume} {92}},\ \bibinfo {pages} {153601} (\bibinfo {year}
  {2004})}\BibitemShut {NoStop}%
\bibitem [{\citenamefont {Ferrini}\ \emph {et~al.}(2013)\citenamefont
  {Ferrini}, \citenamefont {Gazeau}, \citenamefont {Coudreau}, \citenamefont
  {Fabre},\ and\ \citenamefont {Treps}}]{Ferrini2013}%
  \BibitemOpen
  \bibfield  {author} {\bibinfo {author} {\bibfnamefont {G.}~\bibnamefont
  {Ferrini}}, \bibinfo {author} {\bibfnamefont {J.~P.}\ \bibnamefont {Gazeau}},
  \bibinfo {author} {\bibfnamefont {T.}~\bibnamefont {Coudreau}}, \bibinfo
  {author} {\bibfnamefont {C.}~\bibnamefont {Fabre}}, \ and\ \bibinfo {author}
  {\bibfnamefont {N.}~\bibnamefont {Treps}},\ }\href {\doibase
  10.1088/1367-2630/15/9/093015} {\bibfield  {journal} {\bibinfo  {journal}
  {New Journal of Physics}\ }\textbf {\bibinfo {volume} {15}},\ \bibinfo
  {pages} {093015} (\bibinfo {year} {2013})}\BibitemShut {NoStop}%
\bibitem [{\citenamefont {Lamine}\ \emph {et~al.}(2008)\citenamefont {Lamine},
  \citenamefont {Fabre},\ and\ \citenamefont {Treps}}]{Lamine2008}%
  \BibitemOpen
  \bibfield  {author} {\bibinfo {author} {\bibfnamefont {B.}~\bibnamefont
  {Lamine}}, \bibinfo {author} {\bibfnamefont {C.}~\bibnamefont {Fabre}}, \
  and\ \bibinfo {author} {\bibfnamefont {N.}~\bibnamefont {Treps}},\ }\href
  {\doibase 10.1103/PhysRevLett.101.123601} {\bibfield  {journal} {\bibinfo
  {journal} {Physical Review Letters}\ }\textbf {\bibinfo {volume} {101}},\
  \bibinfo {pages} {123601} (\bibinfo {year} {2008})}\BibitemShut {NoStop}%
\bibitem [{\citenamefont {U'Ren}\ \emph {et~al.}(2005)\citenamefont {U'Ren},
  \citenamefont {Silberhorn}, \citenamefont {Erdmann}, \citenamefont
  {Banaszek}, \citenamefont {Grice}, \citenamefont {Walmsley},\ and\
  \citenamefont {Raymer}}]{U'Ren2006}%
  \BibitemOpen
  \bibfield  {author} {\bibinfo {author} {\bibfnamefont {A.~B.}\ \bibnamefont
  {U'Ren}}, \bibinfo {author} {\bibfnamefont {C.}~\bibnamefont {Silberhorn}},
  \bibinfo {author} {\bibfnamefont {R.}~\bibnamefont {Erdmann}}, \bibinfo
  {author} {\bibfnamefont {K.}~\bibnamefont {Banaszek}}, \bibinfo {author}
  {\bibfnamefont {W.~P.}\ \bibnamefont {Grice}}, \bibinfo {author}
  {\bibfnamefont {I.~A.}\ \bibnamefont {Walmsley}}, \ and\ \bibinfo {author}
  {\bibfnamefont {M.~G.}\ \bibnamefont {Raymer}},\ }\href
  {http://arxiv.org/abs/quant-ph/0611019} {\bibfield  {journal} {\bibinfo
  {journal} {Laser Physics}\ }\textbf {\bibinfo {volume} {15}},\ \bibinfo
  {pages} {146} (\bibinfo {year} {2005})}\BibitemShut {NoStop}%
\bibitem [{\citenamefont {Valencia}\ \emph {et~al.}(2007)\citenamefont
  {Valencia}, \citenamefont {Cer\'{e}}, \citenamefont {Shi}, \citenamefont
  {Molina-Terriza},\ and\ \citenamefont {Torres}}]{Valencia2007}%
  \BibitemOpen
  \bibfield  {author} {\bibinfo {author} {\bibfnamefont {A.}~\bibnamefont
  {Valencia}}, \bibinfo {author} {\bibfnamefont {A.}~\bibnamefont {Cer\'{e}}},
  \bibinfo {author} {\bibfnamefont {X.}~\bibnamefont {Shi}}, \bibinfo {author}
  {\bibfnamefont {G.}~\bibnamefont {Molina-Terriza}}, \ and\ \bibinfo {author}
  {\bibfnamefont {J.~P.}\ \bibnamefont {Torres}},\ }\href {\doibase
  10.1103/PhysRevLett.99.243601} {\bibfield  {journal} {\bibinfo  {journal}
  {Physical Review Letters}\ }\textbf {\bibinfo {volume} {99}},\ \bibinfo
  {pages} {243601} (\bibinfo {year} {2007})}\BibitemShut {NoStop}%
\bibitem [{\citenamefont {Gatti}\ \emph {et~al.}(2009)\citenamefont {Gatti},
  \citenamefont {Brambilla}, \citenamefont {Caspani}, \citenamefont
  {Jedrkiewicz},\ and\ \citenamefont {Lugiato}}]{Gatti2009}%
  \BibitemOpen
  \bibfield  {author} {\bibinfo {author} {\bibfnamefont {A.}~\bibnamefont
  {Gatti}}, \bibinfo {author} {\bibfnamefont {E.}~\bibnamefont {Brambilla}},
  \bibinfo {author} {\bibfnamefont {L.}~\bibnamefont {Caspani}}, \bibinfo
  {author} {\bibfnamefont {O.}~\bibnamefont {Jedrkiewicz}}, \ and\ \bibinfo
  {author} {\bibfnamefont {L.~A.}\ \bibnamefont {Lugiato}},\ }\href {\doibase
  10.1103/PhysRevLett.102.223601} {\bibfield  {journal} {\bibinfo  {journal}
  {Physical Review Letters}\ }\textbf {\bibinfo {volume} {102}},\ \bibinfo
  {pages} {223601} (\bibinfo {year} {2009})}\BibitemShut {NoStop}%
\bibitem [{\citenamefont {Law}\ and\ \citenamefont {Eberly}(2004)}]{Law2004}%
  \BibitemOpen
  \bibfield  {author} {\bibinfo {author} {\bibfnamefont {C.~K.}\ \bibnamefont
  {Law}}\ and\ \bibinfo {author} {\bibfnamefont {J.~H.}\ \bibnamefont
  {Eberly}},\ }\href {\doibase 10.1103/PhysRevLett.92.127903} {\bibfield
  {journal} {\bibinfo  {journal} {Physical Review Letters}\ }\textbf {\bibinfo
  {volume} {92}},\ \bibinfo {pages} {127903} (\bibinfo {year}
  {2004})}\BibitemShut {NoStop}%
\bibitem [{\citenamefont {Braunstein}(2005)}]{Braunstein2005}%
  \BibitemOpen
  \bibfield  {author} {\bibinfo {author} {\bibfnamefont {S.~L.}\ \bibnamefont
  {Braunstein}},\ }\href {\doibase 10.1103/PhysRevA.71.055801} {\bibfield
  {journal} {\bibinfo  {journal} {Physical Review A}\ }\textbf {\bibinfo
  {volume} {71}},\ \bibinfo {pages} {055801} (\bibinfo {year}
  {2005})}\BibitemShut {NoStop}%
\bibitem [{\citenamefont {Torres}\ \emph {et~al.}(2010)\citenamefont {Torres},
  \citenamefont {Hendrych},\ and\ \citenamefont {Valencia}}]{Torres2010}%
  \BibitemOpen
  \bibfield  {author} {\bibinfo {author} {\bibfnamefont {J.~P.}\ \bibnamefont
  {Torres}}, \bibinfo {author} {\bibfnamefont {M.}~\bibnamefont {Hendrych}}, \
  and\ \bibinfo {author} {\bibfnamefont {A.}~\bibnamefont {Valencia}},\ }\href
  {\doibase 10.1364/AOP.2.000319} {\bibfield  {journal} {\bibinfo  {journal}
  {Advances in Optics and Photonics}\ }\textbf {\bibinfo {volume} {2}},\
  \bibinfo {pages} {319} (\bibinfo {year} {2010})}\BibitemShut {NoStop}%
\bibitem [{\citenamefont {Torres}\ \emph {et~al.}(2005)\citenamefont {Torres},
  \citenamefont {Maci\`{a}}, \citenamefont {Carrasco},\ and\ \citenamefont
  {Torner}}]{Torres2005a}%
  \BibitemOpen
  \bibfield  {author} {\bibinfo {author} {\bibfnamefont {J.~P.}\ \bibnamefont
  {Torres}}, \bibinfo {author} {\bibfnamefont {F.}~\bibnamefont {Maci\`{a}}},
  \bibinfo {author} {\bibfnamefont {S.}~\bibnamefont {Carrasco}}, \ and\
  \bibinfo {author} {\bibfnamefont {L.}~\bibnamefont {Torner}},\ }\href
  {http://www.ncbi.nlm.nih.gov/pubmed/15751896} {\bibfield  {journal} {\bibinfo
   {journal} {Optics letters}\ }\textbf {\bibinfo {volume} {30}},\ \bibinfo
  {pages} {314} (\bibinfo {year} {2005})}\BibitemShut {NoStop}%
\bibitem [{\citenamefont {Hendrych}\ \emph {et~al.}(2007)\citenamefont
  {Hendrych}, \citenamefont {Micuda},\ and\ \citenamefont
  {Torres}}]{Hendrych2007}%
  \BibitemOpen
  \bibfield  {author} {\bibinfo {author} {\bibfnamefont {M.}~\bibnamefont
  {Hendrych}}, \bibinfo {author} {\bibfnamefont {M.}~\bibnamefont {Micuda}}, \
  and\ \bibinfo {author} {\bibfnamefont {J.}~\bibnamefont {Torres}},\ }\href
  {http://www.opticsinfobase.org/abstract.cfm?\&id=140237} {\bibfield
  {journal} {\bibinfo  {journal} {Optics letters}\ }\textbf {\bibinfo {volume}
  {32}},\ \bibinfo {pages} {2339} (\bibinfo {year} {2007})}\BibitemShut
  {NoStop}%
\bibitem [{\citenamefont {Averchenko}\ \emph {et~al.}(2010)\citenamefont
  {Averchenko}, \citenamefont {Golubev}, \citenamefont {Fabre},\ and\
  \citenamefont {Treps}}]{Averchenko2010}%
  \BibitemOpen
  \bibfield  {author} {\bibinfo {author} {\bibfnamefont {V.}~\bibnamefont
  {Averchenko}}, \bibinfo {author} {\bibfnamefont {Y.}~\bibnamefont {Golubev}},
  \bibinfo {author} {\bibfnamefont {C.}~\bibnamefont {Fabre}}, \ and\ \bibinfo
  {author} {\bibfnamefont {N.}~\bibnamefont {Treps}},\ }\href {\doibase
  10.1140/epjd/e2010-00280-7} {\bibfield  {journal} {\bibinfo  {journal} {The
  European Physical Journal D}\ }\textbf {\bibinfo {volume} {61}},\ \bibinfo
  {pages} {207} (\bibinfo {year} {2010})}\BibitemShut {NoStop}%
\bibitem [{\citenamefont {Jiang}\ \emph {et~al.}(2012)\citenamefont {Jiang},
  \citenamefont {Treps},\ and\ \citenamefont {Fabre}}]{Jiang2012c}%
  \BibitemOpen
  \bibfield  {author} {\bibinfo {author} {\bibfnamefont {S.}~\bibnamefont
  {Jiang}}, \bibinfo {author} {\bibfnamefont {N.}~\bibnamefont {Treps}}, \ and\
  \bibinfo {author} {\bibfnamefont {C.}~\bibnamefont {Fabre}},\ }\href
  {\doibase 10.1088/1367-2630/14/4/043006} {\bibfield  {journal} {\bibinfo
  {journal} {New Journal of Physics}\ }\textbf {\bibinfo {volume} {14}},\
  \bibinfo {pages} {043006} (\bibinfo {year} {2012})}\BibitemShut {NoStop}%
\bibitem [{\citenamefont {Brecht}\ \emph {et~al.}(2014)\citenamefont {Brecht},
  \citenamefont {Eckstein}, \citenamefont {Ricken}, \citenamefont {Quiring},
  \citenamefont {Suche}, \citenamefont {Sansoni},\ and\ \citenamefont
  {Silberhorn}}]{Brecht2014}%
  \BibitemOpen
  \bibfield  {author} {\bibinfo {author} {\bibfnamefont {B.}~\bibnamefont
  {Brecht}}, \bibinfo {author} {\bibfnamefont {A.}~\bibnamefont {Eckstein}},
  \bibinfo {author} {\bibfnamefont {R.}~\bibnamefont {Ricken}}, \bibinfo
  {author} {\bibfnamefont {V.}~\bibnamefont {Quiring}}, \bibinfo {author}
  {\bibfnamefont {H.}~\bibnamefont {Suche}}, \bibinfo {author} {\bibfnamefont
  {L.}~\bibnamefont {Sansoni}}, \ and\ \bibinfo {author} {\bibfnamefont
  {C.}~\bibnamefont {Silberhorn}},\ }\href {http://arxiv.org/abs/1403.4397}
  {\bibfield  {journal} {\bibinfo  {journal} {arXiv:1403.4397v1}\ } (\bibinfo
  {year} {2014})}\BibitemShut {NoStop}%
\bibitem [{\citenamefont {Patera}\ \emph {et~al.}(2009)\citenamefont {Patera},
  \citenamefont {Treps}, \citenamefont {Fabre},\ and\ \citenamefont
  {de~Valc\'{a}rcel}}]{Patera2009}%
  \BibitemOpen
  \bibfield  {author} {\bibinfo {author} {\bibfnamefont {G.}~\bibnamefont
  {Patera}}, \bibinfo {author} {\bibfnamefont {N.}~\bibnamefont {Treps}},
  \bibinfo {author} {\bibfnamefont {C.}~\bibnamefont {Fabre}}, \ and\ \bibinfo
  {author} {\bibfnamefont {G.~J.}\ \bibnamefont {de~Valc\'{a}rcel}},\ }\href
  {\doibase 10.1140/epjd/e2009-00299-9} {\bibfield  {journal} {\bibinfo
  {journal} {The European Physical Journal D}\ }\textbf {\bibinfo {volume}
  {56}},\ \bibinfo {pages} {123} (\bibinfo {year} {2009})}\BibitemShut
  {NoStop}%
\bibitem [{\citenamefont {Osorio}\ \emph {et~al.}(2008)\citenamefont {Osorio},
  \citenamefont {Valencia},\ and\ \citenamefont {Torres}}]{Osorio2008}%
  \BibitemOpen
  \bibfield  {author} {\bibinfo {author} {\bibfnamefont {C.~I.}\ \bibnamefont
  {Osorio}}, \bibinfo {author} {\bibfnamefont {A.}~\bibnamefont {Valencia}}, \
  and\ \bibinfo {author} {\bibfnamefont {J.~P.}\ \bibnamefont {Torres}},\
  }\href {\doibase 10.1088/1367-2630/10/11/113012} {\bibfield  {journal}
  {\bibinfo  {journal} {New Journal of Physics}\ }\textbf {\bibinfo {volume}
  {10}},\ \bibinfo {pages} {113012} (\bibinfo {year} {2008})}\BibitemShut
  {NoStop}%
\end{thebibliography}%

\appendix

\section{Parameters of the nonlinear photon subtraction}\L{Exp}

\subsection{Phase-matching for non-collinear sum-frequency generation}
In the Fig. \ref{fig:non-coll_coord}, signal and gate fields oscillate at equal carrier frequencies $\w_{s,0} = \w_{g,0}$, have ordinary polarizations and are phase-matched for a type-I parametric interaction in an uniaxial non-linear crystal.
We consider the configuration where the input beams and the crystal's optical axis lie in the same plane, coinciding with the plane of the figure.
The up-converted field with a doubled frequency $\w_{c,0}$ and an extraordinary polarization is generated in a phase-matched direction at the angle $\theta$ with respect to the optical axis. When the non-collinear angle $\phi$ is fixed, then the phase-matching condition at carrier frequencies defines the phase-matched angle: $n_e(\w_{g,0}, \theta) = n_o (\w_{s,0}) \cos\phi$, where $n_o, n_e$ are respectively the refraction indices of the ordinary fundamental wave and the extraordinary wave with doubled frequency.
The non-collinear angle is chosen to be small in order to guarantee the spatial overlap of the interacting fields over the entire crystal length.
Also the phase-matching is achieved only for restricted non-collinear angles. The maximal value is given by: $\cos\phi_{max} = n_{e,min}(\w_{g,0})/ n_o (\w_{s,0})$, where $n_{e,min}$ is the minimum value of the index of refraction either at $\theta=0^\circ$ for a positive crystal or at $\theta=90^\circ$ for a negative crystal.
The up-converted field undergoes a birefringent walk-off with an angle $\rho$ in the direction of lower refractive index. Due to the chosen configuration, the beam stays in the plane of the figure.
This walk-off effect brings asymmetry to the non-collinear configuration: exchanging the signal and the gate fields results in a different configuration.

For the numerical estimations of this article, we consider a crystal of BBO, which is uniaxial and negative. The crystal parameters are taken from SNLO software \cite{SNLO}. The crystal is cut to phase-match a type-I non-collinear and degenerate up-conversion process s(o)+g(o)=c(e) at the fundamental wavelength of 800 nm.
The phase-matching can be achieved up to the maximal non-collinear angle $\phi_{max} \approx 19^\circ$.
We consider two angles $\phi=1^\circ$ and $5^\circ$. Phase-matching is then achieved at angles $\theta=29.4^\circ$ and $32.4^\circ$, respectively.
The group velocities of the ordinary polarized signal and gate fields coincide and do not depend on the crystal's cut: $k_s'=k_g'=1.683 \; c^{-1}$, where $c$ is the speed of light in vacuum. For the up-converted field, one obtains $k_c'=1.742\;c^{-1}, \rho=3.9^\circ$ (at $\phi=1^\circ$) and $k_c'=1.735\;c^{-1}, \rho=4.1^\circ$ (at $\phi=5^\circ$).

\subsection{Some properties of multimode squeezed frequency combs}
The spectral profiles of the eigen-modes of a squeezed comb \cite{Roslund2013} are well approximated by Hermite-Gaussian functions with a spectral width $\D \l=6$ nm (FWHM) and centered at 795 nm. In the time domain, a squeezed frequency comb may be represented as a train of quantum correlated pulses of characteristic duration $\t_s = 94$ fs (for the definition, see expression (\ref{a_g})) and a repetition rate of 76 MHz. The number of correlated pulses is roughly equal to the finesse of the optical cavity \cite{Averchenko2010, Jiang2012c} estimated to be ${\cal F} \approx 40$.
Therefore the average number of photons in a single pulse of the given mode may be estimated in the following way
	\begin{align}
	N_{s,n} \approx N_{s,n}^{\text{comb}}/{\cal F} \L{photons-pulse}
	\end{align}
where $N_{s,n}^{\text{comb}}$ is the number of photons in the squeezed mode of the comb.
We calculated the distribution of photons for the experimental parameters by diagonalizing the matrix that describes squeezing of the frequency comb in the optical parametric oscillator \cite{Patera2009}. For this calculation, we bounded the squeezing matrix to 100 nm bandwidth in order to model the spectrally limited finesse of the cavity. This filtering reduces the number of squeezed modes to approximately 40 modes, as seen in Fig. \ref{fig:Ns}.
This figure represents the normalized distribution of photons in different modes. By multiplying it by the number of photons in a pulse of the first mode, one gets the absolute distribution. This yields the estimated multiplication coefficient $N_{s,1} \approx 6\times 10^{-3}$ photons per pulse for the measured 4.2 dB of squeezing of the first mode.

\section{Noncollinear second harmonic generation : calculation of an up-converted state}\L{state}

After the parametric interaction with the classical gate field, the quantum state at the output for the signal and converted fields is given by
	\begin{align}
	& |\textrm{out}\> = {\cal T} \exp\[ \frac{1}{i \hbar} \int\limits \ud t \; \hat H_I(t) \]  |\textrm{in}\>\L{out1}
	\end{align}
where ${\cal T}$ stands for the time-ordering operator. $\hat H_I$ is the Hamiltonian of the parametric process in the interaction picture
	\begin{align}
	& \hat H_I = \varepsilon_0 \int_V \ud V \chi^{(2)} {\hat E}_c^{(-)} {\hat E}_s^{(+)} {E}_g^{(+)} + h.c. \L{H_def}
	\end{align}
where $V$ is the crystal's volume and $\chi^{(2)}$ is an element of the second-order nonlinear susceptibility tensor of the phase-matched process.
In the case of weak parametric up-conversion, the output state (\ref{out1}) may be approximately calculated in the first-order perturbation theory:
	\begin{align}
	|\textrm{out}\> \approx |\textrm{in}\> + |\phi\>
	\end{align}
where
	\begin{align}
	& |\phi\> = \frac{1}{i \hbar} \int\limits \ud t \, \hat H_I(t) \;  |\text{in}\>
	\end{align}
Let us decompose field's amplitudes into plane monochromatic waves.
The decomposition of the up-converted field accounting for the walk-off effect with an angle $\rho$ then reads
	\begin{align}
	\nn & \hat E_c^{(-)}(\xx, z,t) =  {\cal E}_c \int \frac{\ud \w_c \ud \qq_c}{({2 \pi})^{3/2}} \\
	& \times \exp \Big[-i \qq_c (\xx-z \tan\rho) - i k_c z + i \w_c t\Big] \; \hat a_c(\w_c, \qq_c)
	\end{align}
where $\xx = (x,y)$ and $\qq = (q^x,q^y)$ are the transverse coordinates and momenta.
For the signal and the gate fields, it is convenient to define individual coordinate systems such that: $y_{s,g} = y$, $x_{s,g} = x\cos\phi \mp z\sin\phi$ and $z_{s,g} = z\cos\phi \pm y\sin\phi$.
As a result, the signal field may be written as
	\begin{align}
	\nn \hat E_s^{(+)}(\xx_s, & z_s,t) =  {\cal E}_s \int \frac{\ud\w_s \ud\qq_s}{({2 \pi})^{3/2}} \\
	& \times \exp \Big[i \qq_s \xx_s + i k_s z_s - i \w_s t\Big] \; \hat a_s(\w_s, \qq_s)
	\end{align}
The scaling factors ${\cal E}_j$ that are involved in these expressions read: ${\cal E}_{j} = \sqrt{{\hbar \w_{j,0}}/{2 \varepsilon_0 n_{j} c}}$, where $j=c,s$.
The operators obey the commutation relation  $\big[\hat a_j(\w_j), \hat a_j^\dag(\w'_j,\qq'_j)\big] = \dd (\w_j-\w'_j) \dd (\qq_j-\qq'_j)$.
The spatio-spectral profile of the gate field is described by the distribution $\a_g(\w_g, \qq_g)$, normalized as $\int d\w_g d\qq_g |\a_g(\w_g, \qq_g)|^2 =1$. The field's amplitude then reads
	\begin{align}
	\nn E_g^{(+)}(\xx_g, &z_g,t) = \sqrt\frac{W_g}{2\varepsilon_0 n_g c} \int \frac{\ud \w_g \ud \qq_g}{({2 \pi})^{3/2}}\\
	& \times \exp \Big[i \qq_g {\xx}_g + i k_g z_g - i \w_g t\Big] \; \a_g(\w_g, \qq_g)
	\end{align}
where ${W}_g$ is the energy contained in a single pulse of the field.
It is worth to stress that the operators $\hat a_{s,c}$ and the classical quantity $\a_{g}$ correspond to the spectral amplitudes of fields taken at the origin, which we define as being halfway through the crystal (of a total length $l$, as depicted in Fig. \ref{fig:non-coll_coord}). Throughout the analysis, we will neglect the effect of diffraction by assuming that a longitudinal momentum depends only on frequency
	\begin{align}
	& k_j = k_j(\w_j,\qq_j)= \sqrt{k_j^2(\w_j)-|\qq_j|^2} \approx k_j(\w_j)
	%
	\end{align}
where $j=c,s,g$.
Substituting the decomposition of the fields into the Hamiltonian, we perform the integration over the crystal volume, i.e. infinite integration in the transverse direction and integration over the longitudinal interval $z=-l/2 \ldots l/2$. Finally, the integration in time yields the expression for the quantum state (\ref{Hdt}), together with the conservation conditions (\ref{dw})-(\ref{dk}).

\section{Calculation of the Schmidt number for a Gaussian transfer function}\L{Schmidt-calc}
Under the Gaussian approximation, the transfer function (\ref{LWqW}) is conveniently expressed in the form \cite{Osorio2008}
	\begin{align}
	& L(\W_c,q_c, \W_s) = {\cal C} \exp\[-\frac{1}{2} x^T U x\]
	\end{align}
where $x^T = (\W_c,q_c, \W_s)$, $U$ is a $3\times 3$ covariance matrix and ${\cal C}$ is a proportionality coefficient.
The Schmidt decomposition of the function into signal and up-converted parts is expressed as
	\begin{align}
	& L(\W_c,q_c, \W_s) = \sum\limits_{m=1}^{\infty} \l_m \; \psi_m(\W_c,q_c) \; \varphi_m^*(\W_s) \L{decomp}
	\end{align}
The definition of the Schmidt number reads
	\begin{align}
	& K=\frac{\big(\sum\limits_m \l_m^2\big)^2}{\sum\limits_m \l_m^4}\L{K_3}
	\end{align}
Now to calculate the Schmidt number, let us note that the sum of the squared coefficients $\sum\limits_{m} \l_m^2$ is equal to the norm of the transfer function
	\begin{align}
	\nn\sum\limits_{m} \l_m^2 &= \int \ud\W_c \ud q_c \ud\W_s \left|L(\W_c,q_c, \W_s)\right|^2 \\&= {\cal C}^2 \frac{(2\pi)^{3/2}}{\sqrt{\text{Det}(2U)}} \L{l2}
	\end{align}
To perform this integration, we used the general expression for multivariate Gaussian integrals
	\begin{align}
	\idotsint \exp\[-\frac{1}{2}q^T M q\] \ud q_1 \ldots \ud q_n = \frac{(2\pi)^{n/2}}{\sqrt{\text{Det}(M)}}
	\end{align}
To calculate $\sum\limits_{m} \l_m^4$, we note that
	\begin{align}
	\nn G(\W_s,\W_s') &\equiv \int \ud{\W}_c \ud {q}_c \; L^*({\W}_c,{q}_c, \W_s) L({\W}_c,{q}_c, \W'_s) \\
	&= \sum\limits_m \l_m^2 \; \varphi_m(\W_s) \; \varphi_m^*(\W'_s)
	\end{align}
where we used the decomposition (\ref{decomp}). Calculation of the norm of the auxiliary function then gives
	\begin{align}
	\nn \sum\limits_{m} \l_m^4 & = \int \ud{\W}_s \ud {\W}_s' \left|G(\W_s,\W_s')\right|^2\\
	& = {\cal C}^4 \frac{(2\pi)^{6/2}}{\sqrt{\text{Det}(V)}} \L{l4}
	\end{align}
where $V$ is a $6\times 6$ covariance matrix defined by
	\begin{align}
	\nn & {\cal C}^4 \exp\[-\frac{1}{2} X^T V X\] \\
	& = L^*({\W}_c,{q}_c, \W_s) L({\W}_c,{q}_c, \W'_s) L(\W'_c,q'_c, \W_s) L^*(\W'_c,q'_c, \W'_s)
	\end{align}
with $X^T=\big(x^T,x'^T\big)=\big(\W_c,q_c, \W_s,\W'_c,q'_c, \W'_s\big)$.
Substituting (\ref{l2}) and (\ref{l4}) into (\ref{K_3}), one gets the expression of the Schmidt number through covariance matrices
	\begin{align}
	K = \frac{\sqrt{\text{Det}(V)}}{{\text{Det}(2 U)}}
	\end{align}
\end{document}